\documentclass[fleqn,usenatbib]{mnras}

\usepackage{newtxtext,newtxmath,pdflscape}
\usepackage{siunitx}

\usepackage[T1]{fontenc}

\DeclareRobustCommand{\VAN}[3]{#2}
\let\VANthebibliography\thebibliography
\def\thebibliography{\DeclareRobustCommand{\VAN}[3]{##3}\VANthebibliography}

\usepackage{listings}
\usepackage{color} 
\usepackage{svg} 
\usepackage{fontawesome} 

\definecolor{codegreen}{rgb}{0,0.6,0}
\definecolor{codegray}{rgb}{0.5,0.5,0.5}
\definecolor{codepurple}{rgb}{0.58,0,0.82}
\definecolor{backcolour}{rgb}{0.95,0.95,0.92}

\usepackage{graphicx}	
\usepackage{amsmath}	

\usepackage[dvipsnames]{xcolor} 

\newcommand{\spice}{\texttt{SPICE}}
\newcommand{\spips}{{\tt SPIPS}}
\newcommand{\phoebe}{{\tt PHOEBE}}

\newcommand{\jax}{{\tt JAX}}
\newcommand{\numpy}{{\tt NumPy}}
\newcommand{\tpayne}{{\tt TransformerPayne}}

\title[SPICE]{SPICE\thanks{\url{https://github.com/maja-jablonska/spice}} - modelling synthetic spectra
of stars with non-homogeneous surfaces}

\author[M. Jab{\l}o\'nska et al.]{M. Jab{\l}o\'nska,$^{1}$\thanks{E-mail: maja.jablonska@anu.edu.au}
T. R\'o\.za\'nski,$^{1,2}$
Luca Casagrande,$^{1}$
Hilay Shah$^{1}$,
P. A. Ko{\l}aczek-Szyma\'nski,$^{2,3}$
M. Rychlicki,$^{4}$ \newauthor
and Yuan-Sen Ting$^{5,6}$
\\
$^{1}$Research School of Astronomy \& Astrophysics, Australian National University, Cotter Rd., Weston, ACT 2611, Australia\\
$^{2}$Astronomical Institute, University of Wroc\l aw, Kopernika 11, 51-622 Wroc\l aw, Poland\\
$^{3}$Space Sciences, Technologies and Astrophysics Research (STAR) Institute, Université de Liège, Quartier Agora, Allée du 6 Août 19c, Bât. B5c, 4000, Liège, Belgium\\
$^{4}$School of Computing, University of Leeds, Leeds, UK\\
$^{5}$School of Computing, Australian National University, Acton, ACT 2601, Australia\\
$^{6}$Department of Astronomy, The Ohio State University, Columbus, USA\\
}

\date{Accepted XXX. Received YYY; in original form ZZZ}

\pubyear{\the\year{}}

\begin{document}
\label{firstpage}
\pagerange{\pageref{firstpage}--\pageref{lastpage}}
\maketitle

\begin{abstract}
In the era of large time-domain spectro-photometric surveys, surface variations such as starspots, chemical inhomogeneities, pulsations, rotational distortions, and binary interactions can now be directly detected and modelled. Accurately interpreting these phenomena requires stellar spectral synthesis frameworks that go beyond the assumption of homogeneous surface properties. Yet most existing tools remain limited by this simplification, hindering their applicability to stars with complex surface structures. To address this need, we present \spice\ (SPectral Integration Compiled Engine), an open-source Python package for generating high-resolution spectra and photometry from non-homogeneous stellar surface models. 
\spice\ integrates angle-dependent specific intensities from each surface element, enabling forward modelling of both photometric and spectroscopic variability. 
Case studies demonstrate applications to spotted stars, Cepheid pulsations, and eclipsing binaries, making \spice\ well-suited for analysing current and upcoming survey data. In addition, \spice\ can directly import meshes from \phoebe, enabling the modelling of complex binary configurations beyond these case studies.
\end{abstract}




\section{Introduction}

Synthetic fluxes used to infer stellar properties are typically generated with one‑dimensional, plane‑parallel or spherically symmetric model atmospheres \citep[e.g.,][]{Sneden_1973,Kurucz,Baron_Chen_Hauschildt_2010, Plez_2012, Wheeler_Abruzzo_Casey_Ness_2023}. Although modern three-dimensional simulations capture small-scale convective inhomogeneities \citep[e.g.,][]{freytag12, magic14} they still assume globally homogeneous surfaces when computing emergent fluxes.\footnote{A subtlety here is that in 3D models, the input set of stellar parameters is homogeneous, but convection induces temperature variations. Importantly, temperature itself is not an input parameter - only entropy is.} This assumption simplifies modelling but does not reflect many physically relevant scenarios.

Stars exhibit a wide variety of surface structures and asymmetries, including temperature or abundance spots, chemical patches, rotational flattening with associated gravity darkening, and surface distortions caused by pulsations. In binary systems, additional effects—such as reflection, eclipses, and tidal interactions—modify the observed flux. These features introduce wavelength- and time-dependent variability that homogeneous surface models cannot capture.

Interferometric, photometric and spectroscopic observations can help detect or infer some of these features \citep[e.g.,][]{achernar,delta_persei,cordoni,turisgallo2025unveilingstellarspindetermining}. In this context, time-series observations are particularly powerful for probing stellar surface inhomogeneities. Doppler Imaging, which reconstructs the properties of unresolved stellar surfaces using high-resolution spectroscopic data with adequate time coverage \citep{Vogt_Penrod_1983,Marsh_Horne_1988,Berdyugina,Kochukhov_2016}, is a prominent technique; however, it is demanding both in terms of modelling complexity and observational requirements. In addition, the commonly used $\chi^2$ fitting method is often susceptible to degeneracies \citep{Klein_2025}.

The growing volume of spectroscopic and photometric data from large surveys further motivates the development of more realistic forward models. Contemporary surveys routinely provide hundreds of thousands to millions of spectra with derived stellar parameters. Examples include APOGEE \citep{Majewski_Schiavon_Frinchaboy_Allende_Prieto_Barkhouser_Bizyaev_Blank_Brunner_Burton_Carrera_et_al._2017}, GALAH \citep{De_Silva_Freeman_Bland-Hawthorn_Martell_De_Boer_Asplund_Keller_Sharma_Zucker_Zwitter_et_al._2015, Buder_Sharma_Kos_Amarsi_Nordlander_Lind_Martell_Asplund_Bland-Hawthorn_Casey_et_al._2021}, LAMOST \citep{Cui_Zhao_Chu_Li_Li_Zhang_Su_Yao_Wang_Xing_et_al._2012}, RAVE \cite{Steinmetz_2006}, and \textit{Gaia}-ESO \citep{Gilmore_Randich_Asplund_Binney_Bonifacio_Drew_Feltzing_Ferguson_Jeffries_Micela_et_al._2012, Randich_Gilmore_Magrini_Sacco_Jackson_Jeffries_Worley_Hourihane_Gonneau_Vàzquez_et_al._2022}, with upcoming surveys such as 4MOST \citep{4MOST} or WEAVE \citep{WEAVE} expanding this dataset further. The latest \textit{Gaia} data release provides low-resolution BP/RP spectrophotometry for about two hundred million targets \citep{De_Angeli_Weiler_Montegriffo_Evans_Riello_Andrae_Carrasco_Busso_Burgess_Cacciari_et_al._2023}, and higher resolution (R$\simeq$ 11500) spectra of the calcium triplet for about 1 million sources \citep{gaia_rvs}, with the upcoming \textit{Gaia} DR4 making individual time series spectra also available. Soon, the Vera C. Rubin Observatory's Legacy Survey of Space and Time (LSST) will revolutionise time-domain astronomy by scanning the entire southern sky every few nights, generating an unprecedented stream of time-resolved photometric data for billions of objects. This will open new windows into stellar variability, transient phenomena, and the dynamic behaviour of stars across the Hertzsprung–Russell diagram. Forward-modelling via simulated stars and synthetic spectra with certain parameters, rather than backwards-modelling from stellar spectra to infer parameters, will become increasingly important as newer, high-resolution data are obtained from these ongoing surveys.

Several existing tools address individual aspects of surface inhomogeneities or binary modelling. \phoebe\ provides sophisticated mesh-based modelling of stellar surfaces and is particularly powerful for binary star systems, but focuses primarily on photometric modelling with limited high-resolution spectroscopic capabilities \citep{phoebe}. The spectroscopic module, currently under development, will extend \phoebe\ by enabling the generation of synthetic spectra through interpolation over precomputed spectral grids, assuming a geometric limb-darkening law \citep{phoebe_spectro}. \texttt{CoMBiSpeC} adopts a similar strategy, interpolating over spectral grids while applying an analytic limb-darkening law and accounting for key effects in massive binary systems, including tidal interactions, reflection, and radiation pressure \citep{combispec}. \texttt{SPAMMS} builds on \texttt{PHOEBE}’s 3D mesh models to generate spectra of massive stars, operating on emergent intensities rather than fluxes and incorporating additional effects such as stellar winds \citep{spamms}. \texttt{ZPKETR} effectively captures the effects of rapid rotation on stellar spectra \citep{zpektr}. \texttt{ESTER} offers sophisticated two-dimensional modelling of rotating stellar structures with self-consistent determination of differential rotation and meridional circulation, but is designed primarily for internal structure modelling rather than detailed spectroscopic synthesis \citep{ESTER}. \texttt{STARRY} employs a probabilistic approach to infer the spot populations from light curves \citep{Luger_2019}. \texttt{SOAP-GPU} offers faculae and spot modelling on a spectral level using the \texttt{PHOENIX} library \citep{soapgpu, phoenix}. 

Building upon the capabilities of existing tools, we introduce a unified software package that extends their scope by enabling efficient and flexible modelling of a broad range of stellar surface phenomena across many spectral types, at the scale required by modern surveys. \spice\ presented here fills this gap by providing an open-source, modular Python package capable of simulating non-homogeneous stellar surfaces, including effects such as pulsations, rotation, star spots, and eclipsing binaries, within a single, coherent architecture. \spice\ allows users to define both the surface mesh and the grid of inferred parameters, as well as to customise the underlying spectral synthesis model, making it broadly extensible to diverse physical processes and stellar environments while maintaining computational efficiency for large-scale applications.

The structure of this paper is as follows. In 
Section \ref{sec:methods}, we outline the specifics of the stellar mesh model adopted in \spice. Section \ref{sec:results} focuses on qualitative checks and demonstrations of the current version of the software. Section \ref{sec:conclusions} summarises the paper and outlines future work.

\section{Methods}
\label{sec:methods}

\spice\ uses triangular mesh-based stellar surface models to calculate synthetic spectra as the sum of individual surface element contributions to the total flux. With this approach, we can account for various inhomogeneities in stellar parameters, including temperature spots, pulsations, chemical spots, rotational distortions, and binary interactions. The software leverages automatic differentiation capabilities provided by the \jax\ package, enabling efficient computation of spectrum flux gradients with respect to input parameters, such as effective temperature or rotation velocity. The mesh-based approach also enables seamless integration with photometric modelling software, such as \phoebe, allowing for consistent modelling of both photometric and spectroscopic time series of binary systems.

\begin{figure*}
\centering
    \includegraphics[width=\textwidth]{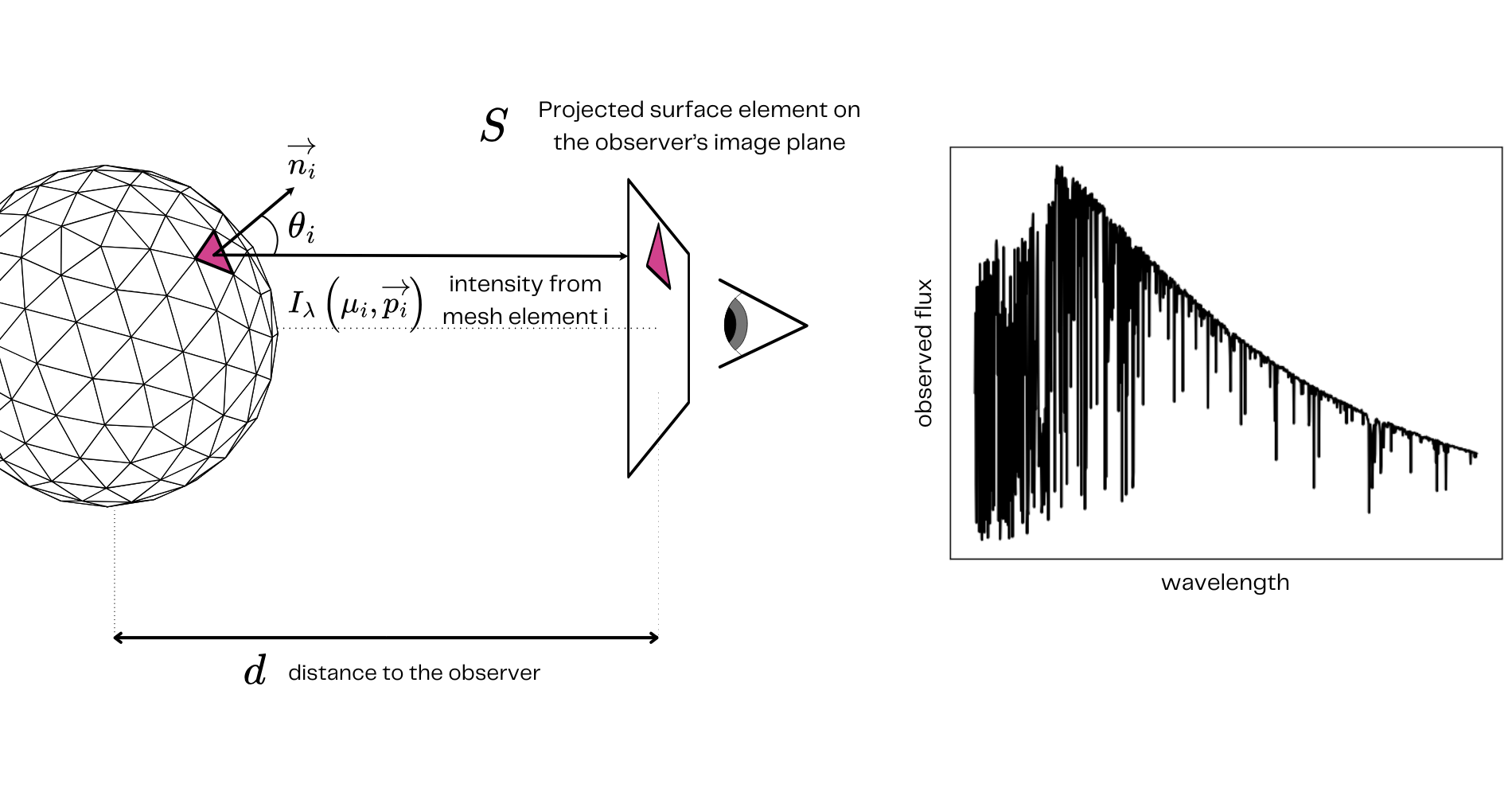}
    \caption{Synthetic observed-flux computation model for a triangulated stellar surface. The left panel illustrates the geometry of the synthetic mesh, where each triangular element $i$ (highlighted in pink) contributes an intensity $I_{\lambda,i}$ that is modulated by the cosine of the angle $\mu_i = \cos\theta_i$ between the surface normal vector $\vec{n}_i$ and the observer’s line of sight. The projection of the surface element with area $S$ onto the observer’s plane is also indicated. \spice\ assumes parallel light rays (i.e. the star is treated as a point source at infinite distance), which is appropriate for all but the nearest resolved stellar surfaces. The right panel shows an example of the resulting synthetic flux $F_\lambda$, obtained by summing the contributions from all visible mesh elements.}
    \label{fig:spice_main_plot}
\end{figure*}

\begin{figure}
    \centering
    \includegraphics[width=\columnwidth]{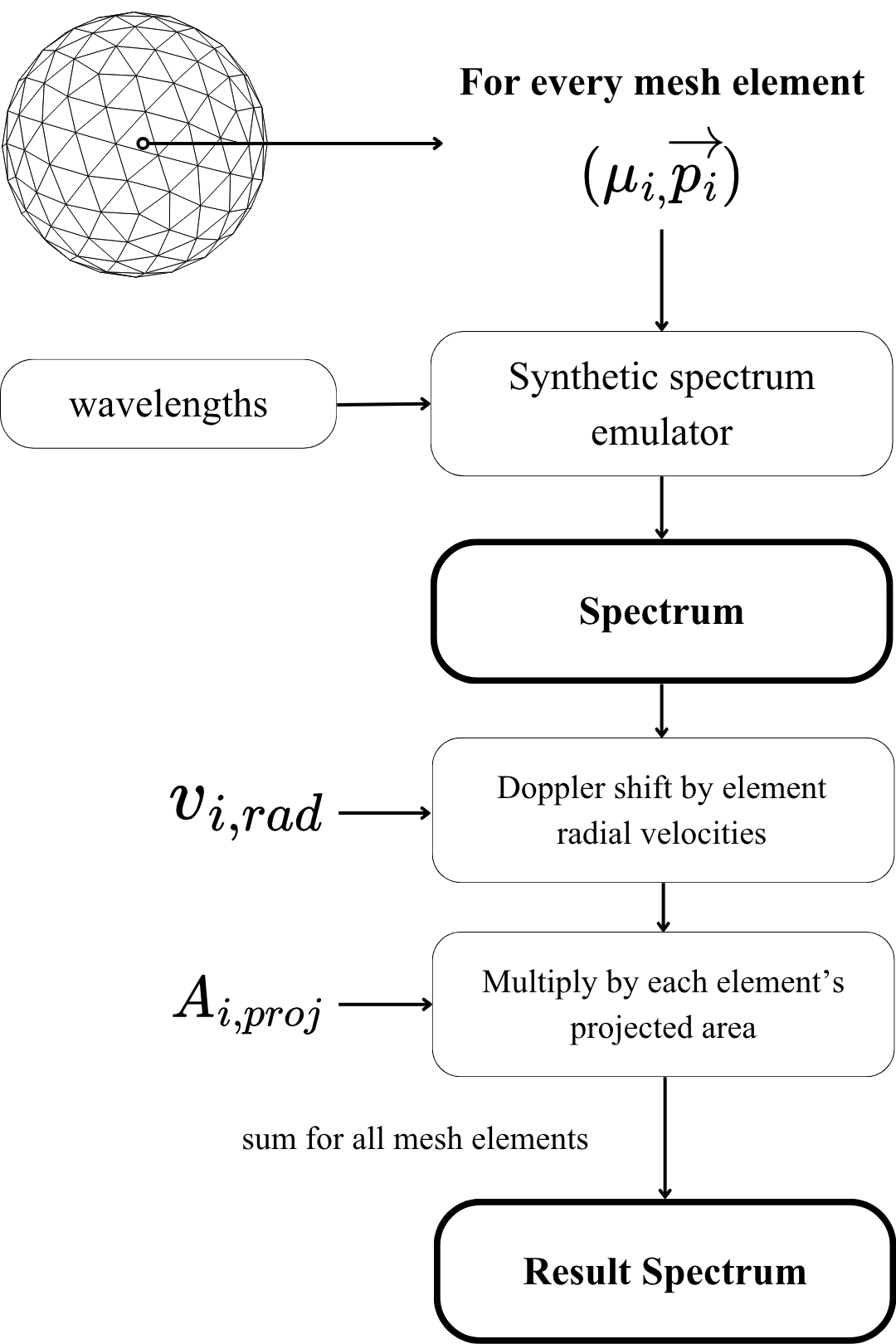}
    \caption{Schematic overview of the spectrum-synthesis workflow in \spice. The diagram summarises the main steps from mesh construction and surface-parameter assignment, through spectral-intensity emulation, to the integration of synthetic spectra.}
    \label{fig:spice_diagram}
\end{figure}

\subsection{Tessellation}

To achieve a flexible, efficient, and parallelisable framework for modelling stellar surfaces at high fidelity, we represent the star with a 3D triangular mesh that captures surface inhomogeneities (e.g.\ spots) and physical distortions (e.g.\ pulsations or rotational flattening).

The mesh is a discretised approximation of a sphere. Each surface element is defined by three vertex position vectors $\vec{r}_{i,j}$ ($j=1,2,3$) and a velocity vector $\vec{v}_i$ for its centre. From these, we compute the intrinsic area $A_i$ and assign a vector of local atmospheric parameters $\vec{p}_i$ (e.g.\ $T_{\rm eff}$, $\log g$, abundances), which are intrinsic to the star and independent of the observer.

An effective tessellation requires uniform element distribution, consistent element sizes, and computational efficiency. We therefore adopt an icosphere discretisation—a widely used method in computer graphics—that yields nearly uniform triangular elements with minimal area variation. This uniformity ensures consistent treatment of surface features across all latitudes, avoiding the distortions inherent in latitude–longitude grids. Because each triangle can be processed independently, the discretisation also enables efficient parallelisation of flux and intensity calculations.

The mesh resolution can be tailored to the problem: coarse grids for global phenomena (e.g.\ rotation or pulsation) and finer grids for local structures (e.g.\ spots or chemical patches), balancing accuracy and computational cost. The overall flux computation procedure is illustrated in Figure~\ref{fig:spice_main_plot}.

The stellar surface is implemented as an icosphere formed by subdividing the faces of a regular icosahedron and projecting the new vertices onto the sphere. Each subdivision level quadruples the number of triangular elements, producing a set of discrete mesh resolutions. The software automatically rounds the user’s requested element count to the nearest valid subdivision level. Detailed implementation notes and comparisons between the icosphere geometry and a perfect sphere are provided in Appendix~\ref{sec:appendix_mesh_construction}.

\subsection{Integrating flux from mesh representing stellar surface}
\label{sec:integrating_flux_from_mesh}

To integrate the flux, one needs to specify the direction to the observer $\hat{n}_{\text{obs}}$ and the distance between the centre of a stellar mesh and the observer $d$. Once the observer position is defined, we can determine the view-dependent quantities:
$$
\mu_i = \hat{n}_{\text{obs}}\!\cdot\!\hat{n}_i,
\qquad
v_{{\rm rad},i} = \vec{v}_i\!\cdot\!\hat{n}_{\text{obs}},
$$
where $\mu_i = \hat{n}_{\text{obs}}\cdot\hat{n}_i = cos\ \theta$, and $\hat{n}_i$ and $\vec{v}_i$ are the facet’s outward unit normal and mean velocity, respectively. The resulting radial velocity $v_{{\rm rad},i}$ naturally accounts for effects such as rotational broadening, pulsations, and orbital motion. 
The radial velocity Doppler‑shifts a predefined wavelength grid (we define the receding velocity as negative): $\vec{\lambda}$ to
$$
\vec{\lambda}_{{\rm d},i} = -\vec{\lambda}\bigl(1 + v_{{\rm rad},i}/c\bigr).
$$

We characterize each mesh element by the set $(A_i, \mu_i, v_{{\rm rad},i}, \vec{p}_i)$, where $A_i$ is the surface area, calculated as the fractional contribution (resulting from the icosphere construction) of the total surface area determined from the provided stellar radius, and $\vec{p}_i$ is a set of physical parameters.

For each mesh element, we use a wavelength-wise intensity emulator to compute specific intensity at the $\mu_i, p_i$ of the mesh and radial velocity shift it:
$I_{\lambda_{{\rm d},i}}(\mu_i, \vec{p}_i)$ for every wavelength in $\vec{\lambda}_{{\rm d},i}$ \footnote{The specifics of the intensity emulator itself are discussed in Section~\ref{sec:spectrum_emulator}.}. 
We calculate the monochromatic observed flux for a specific observer at their position by integrating the contributions from all visible surface elements as 
\begin{equation}
    f_\lambda = \sum_{i=1}^{N_{\rm visible}} I_{\lambda_{\text{d}, i}}(\mu_i, \vec{p}_i) \cdot A_{\mathrm{proj},i} \cdot \frac{1}{d^2},
\label{eq:synthetic_flux}
\end{equation}

where $A_{\mathrm{proj},i} = \mu_i A_i$ denotes the projected visible surface area and $d$ is the distance to the observer. Summing the contributions from all visible surface elements yields the total observed stellar flux, which scales inversely with the square of the distance, following the $1/d^2$ dependence. Accurately handling the viewing angle $\mu_i$ is a critical component of our approach, as it is essential to incorporate relevant atmospheric effects, particularly limb darkening, which directly impacts the precision of spectrum synthesis using the intensity emulator. We outline the complete workflow for the spectrum generation procedure in Figure \ref{fig:spice_diagram}.

Our model operates under the assumption of parallel light rays, an approximation that remains valid for most stellar applications, given that stellar angular sizes are negligible from observational distances. While the distance parameter may not be required for all applications, we include it here to maintain theoretical completeness and formalism. Further details on the geometric quantities entering SPICE can be found in Appendix \ref{sec:appendix_radiative_quantities}.

\subsection{Stellar Surface Properties}

The current version of \spice\ makes it possible to model the following stellar surface phenomena:
\begin{itemize}
    \item \textbf{rigid rotation}
    \item \textbf{spots}: defined either by coordinates and radii, or using spherical harmonics functions,
    \item \textbf{pulsations}: characterised by spherical harmonics functions and Fourier series coefficients.
\end{itemize}
\spice\ also supports the computation of spectra for two independent meshes, where stellar occlusions are resolved at the individual mesh triangle level, enabling detailed modelling of the complex binary star phenomena. This method is described in more detail in Section \ref{sec:binary_systems}.

\begin{table}
    \caption{Number of mesh elements for the corresponding subdivision levels.}
    \label{tab:subdivs_mesh_triangles}
    \begin{tabular}{cc}
        \hline
        \textbf{Number of subdivisions} & \textbf{Number of mesh elements} \\
        \hline
        1 & 80 \\
        2 & 320 \\
        3 & 1280 \\
        4 & 5120 \\
        5 & 20480 \\
        \hline
    \end{tabular}
\end{table}

\subsubsection{Rotation}

\begin{figure}
    \centering
    \includegraphics[width=\linewidth]{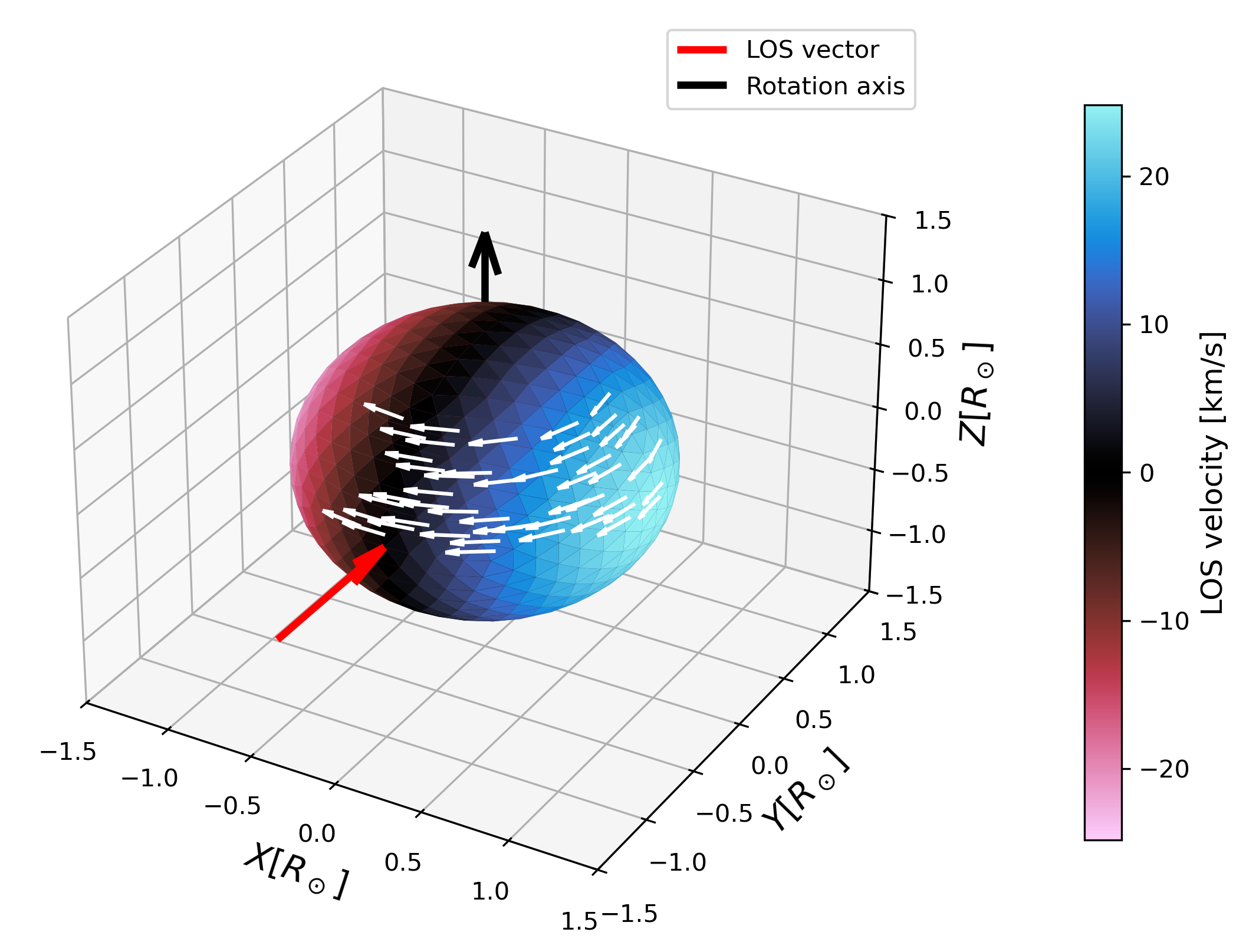}
    \caption{Three-dimensional velocity field of a rotating stellar model. White arrows show velocity vectors for a representative subset of mesh elements for clarity. The colour scale indicates line-of-sight (LOS) velocities computed from the full 3D velocity field and a chosen viewing direction. In \spice, receding velocities are defined as negative.}
    \label{fig:velocity_vectors}
\end{figure}

The current version of the package only supports rigid rotation. We specify the rotation using the equatorial velocity, \(v_{\mathrm{rot}}\), and the normalised rotation axis, \(\hat{n}_{\omega}\), in the star's reference frame. From this axis, we construct the skew-symmetric matrix
\begin{equation}
\mathcal{R} = \begin{bmatrix}
0 & -\hat{n}_{\omega,z} & \hat{n}_{\omega,y} \\
\hat{n}_{\omega,z} & 0 & -\hat{n}_{\omega,x} \\
-\hat{n}_{\omega,y} & \hat{n}_{\omega,x} & 0
\end{bmatrix},
\label{eq:skew_matrix}
\end{equation}
which is used in Rodrigues' rotation formula \citep{rodrigues1840} to generate the rotation matrix. At each time step, we evaluate the rotation matrix ($\mathcal{R}$) and its derivative ($\mathcal{R}'$) to update the coordinates and velocities of the mesh elements. This yields a velocity field with three-component vectors assigned to each element, as illustrated in Figure~\ref{fig:velocity_vectors}. For clarity, we show only a subset of vectors. We can then directly extract line-of-sight velocities for any specified viewing direction.

\subsubsection{Spots}

\begin{figure}
    \centering
    \includegraphics[width=\linewidth]{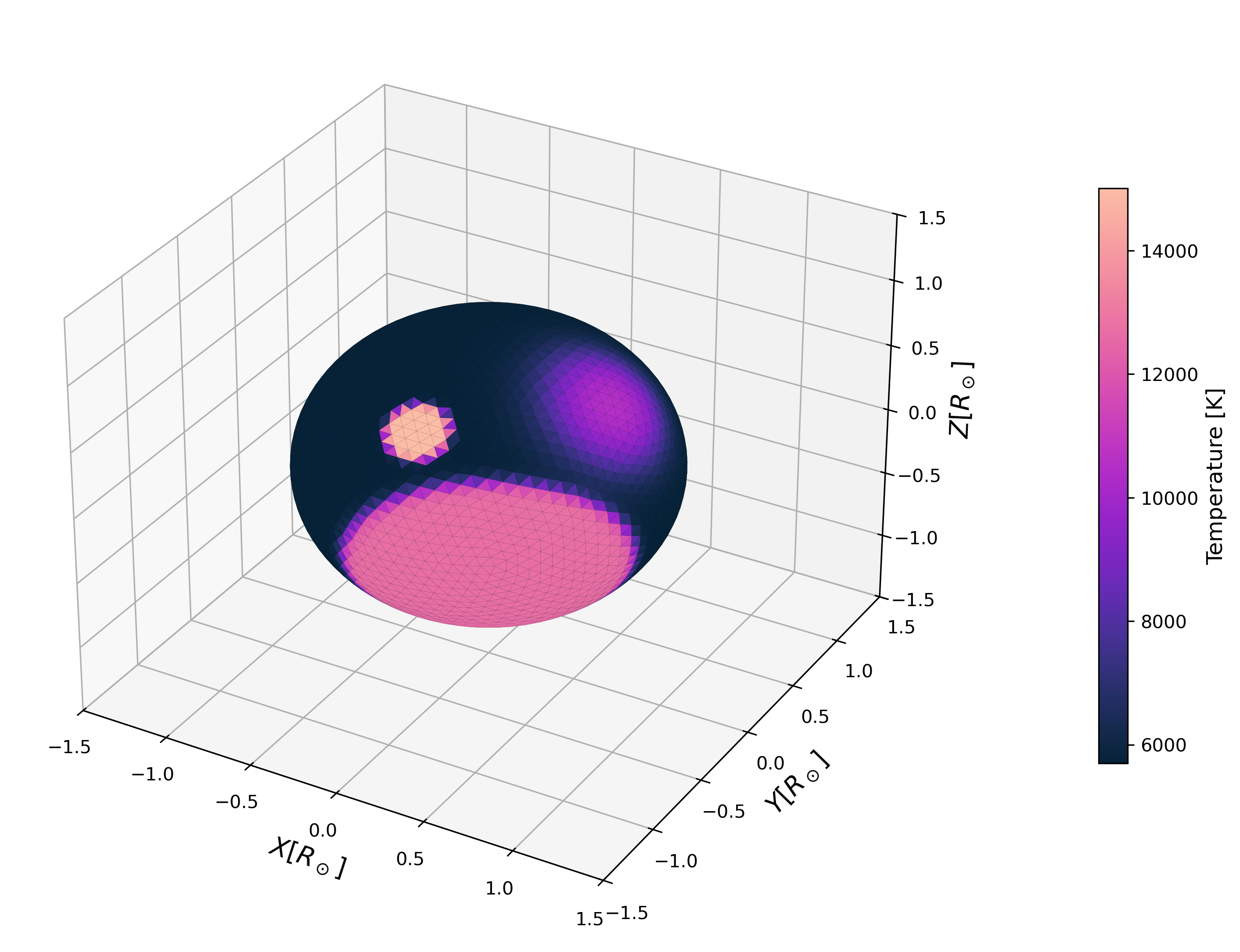}
    \caption{Example of a stellar surface with spots of different sizes and smoothness parameters. The illustration highlights how varying the spot radius and smoothness parameter $\alpha$ leads to sharper or more gradual transitions between the spot and the surrounding surface.}
    \label{fig:spot_examples}
\end{figure}

\begin{figure}
    \centering
    \includegraphics[width=\linewidth]{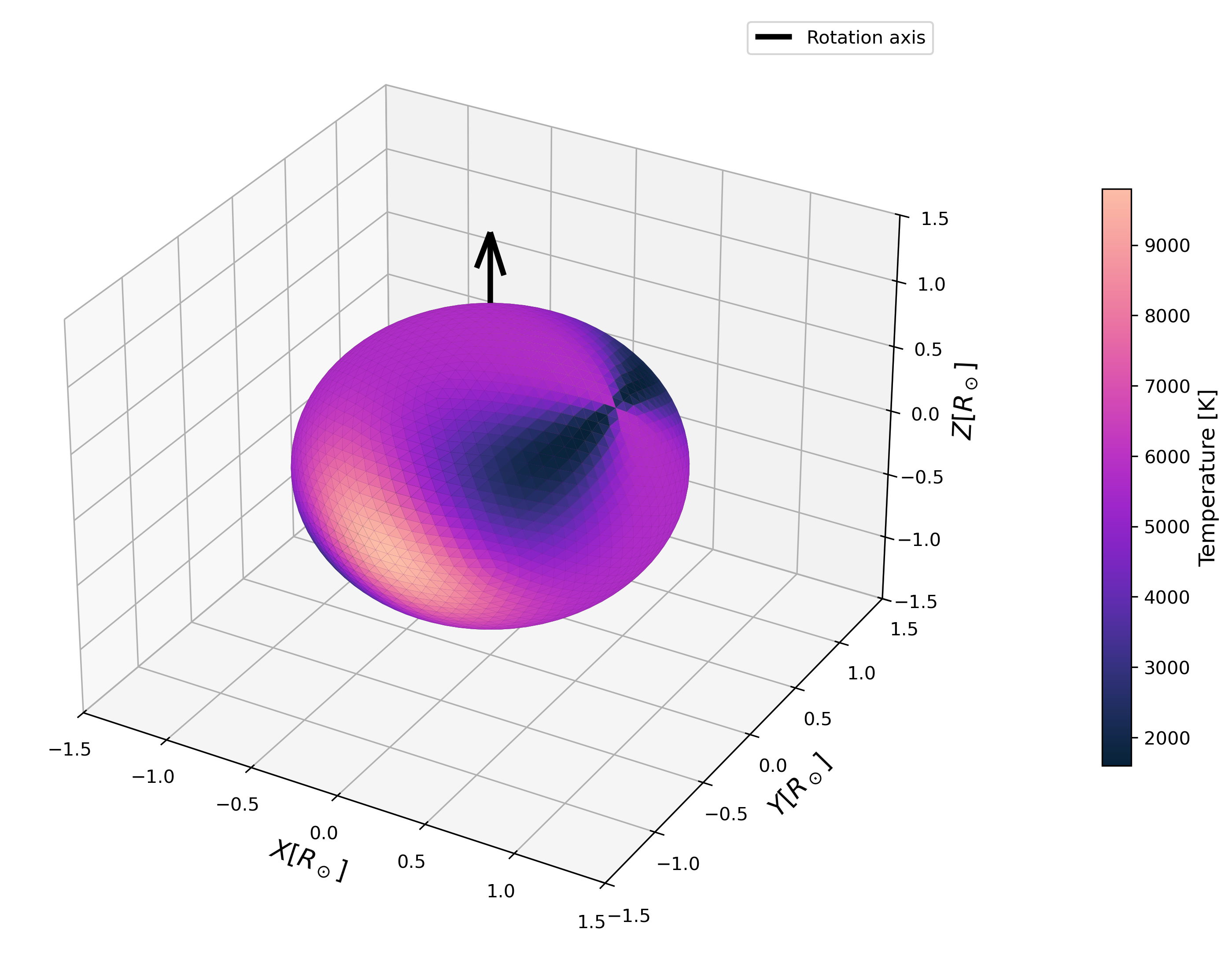}
    \caption{Example stellar surface with a temperature distribution defined by a spherical harmonic function with $m = 4$ and $\ell = 4$, tilted by $45^\circ$ with respect to the rotation axis. This configuration illustrates how spherical harmonics can capture structured, non-axisymmetric surface features.}
    \label{fig:sph_ham_spot_example}
\end{figure}

Spots can be defined for any parameter (most commonly temperature and abundance) using two different approaches:

\begin{enumerate}
    \item By specifying spot latitude and longitude ($\phi$, $\theta$), the spot radius in degrees, parameter value difference $\Delta p_\text{max}$, and smoothness parameter $\alpha$. The smoothness parameter controls the transition between the spot centre and the background value using a rescaled \texttt{sigmoid} function. The smoothness function is demonstrated in detail in Appendix \ref{sec:appendix_spots_smoothness}. An example of a spot configuration with spots with various parameter values is shown in Figure \ref{fig:spot_examples}.

    \item Alternatively, spots can be defined using spherical harmonic functions, which is likely more convenient for inference purposes. These orthogonal functions are defined on a sphere's surface using the parameters of spherical degree $\ell$ and azimuthal order $m$. In \spice, only the real components are utilised, and spots are characterised by parameters $\ell$, $m$, and the maximum parameter difference, which scales the calculated harmonic sphere function values. An example of a spot configuration defined by a harmonic function is shown in Figure \ref{fig:sph_ham_spot_example}.
\end{enumerate}

\subsubsection{Pulsations}

\begin{figure*}
    \centering
    \includegraphics[width=\linewidth]{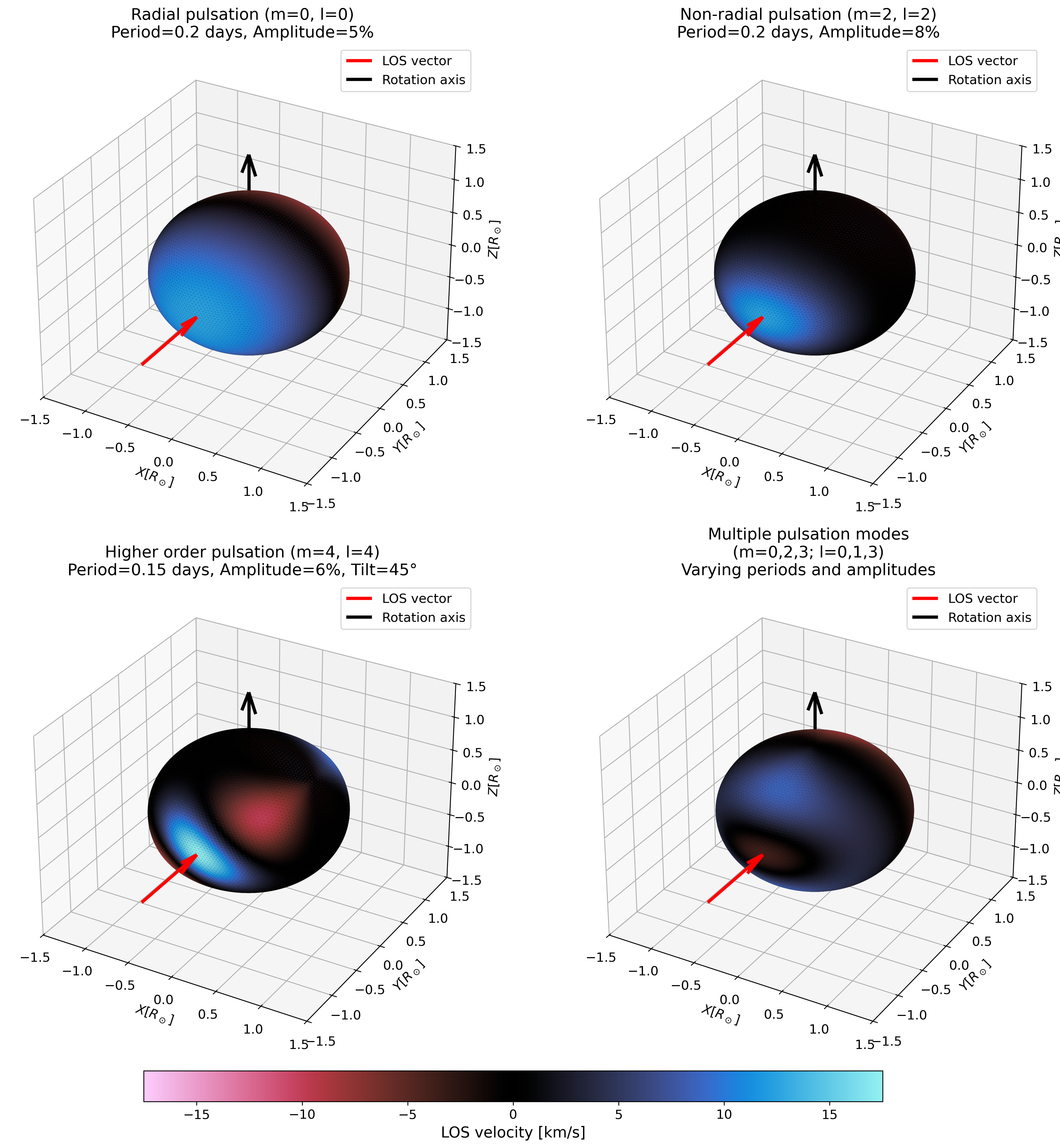}
    \caption{Examples of pulsation patterns defined by spherical harmonic functions evaluated at $t=0.05$ d. Each panel displays a distinct combination of spherical degree $\ell$ and azimuthal order $m$, showcasing the range of spatial structures that can be modelled with the pulsation formalism implemented in \spice.}
    \label{fig:pulsations_examples}
\end{figure*}

\spice\ models pulsations as temporally varying radial displacements across the stellar surface. Spatial variations are described by spherical harmonics $Y^m_\ell(\theta,\phi)$, which are rescaled to $(-1,1)$ to ensure consistent amplitude scaling across modes. Fourier series represent temporal variations.

The radial displacement at time $t$ and position $(\theta,\phi)$ is
\begin{equation}
A(t,\theta,\phi)=Y^m_\ell(\theta,\phi)
\sum_{n=1}^{N}D_n \cos\!\left(\frac{2\pi n}{P}t - \phi_n\right),
\end{equation}
where $D_n$ and $\phi_n$ are Fourier amplitudes and phases, and $P$ is the pulsation period. The corresponding radial velocity is the time derivative:
\begin{equation}
A'(t,\theta,\phi)=Y^m_\ell(\theta,\phi)
\sum_{n=1}^{N}
\left(
-\frac{2\pi n}{P} D_n \sin\!\left(\frac{2\pi n}{P}t - \phi_n\right)
\right).
\end{equation}

Radial pulsations correspond to the spherically symmetric case with $\ell = m = 0$. More complex pulsation patterns are obtained by incorporating higher-order spherical harmonics $Y_{\ell}^{m}$ with $\ell > 0$ or $m \neq 0$, yielding non-radial modes with spatially varying displacement patterns across the stellar surface. Overtone modes, representing higher-frequency harmonics of the fundamental oscillation, are introduced by adding multiple Fourier components with distinct amplitudes, frequencies, and phase offsets to the decomposition of $A(t,\theta,\phi)$. Together, these ingredients enable \spice\ to model a broad spectrum of pulsation phenomena, ranging from simple radial oscillations to complex multi-mode patterns that combine several angular and temporal components.

\subsubsection{Binary systems}
\label{sec:binary_systems}

Beyond modelling a single stellar surface, SPICE can also combine two meshes to simulate binary systems. The binarity option evolves both bodies along their Keplerian orbits and computes their mutual occlusions. To determine the visibility of each mesh fragment, we employ the Sutherland--Hodgman algorithm, which calculates intersection areas between pairs of mesh elements. To improve computational efficiency during this process, a KD-tree-based spatial partitioning structure is used to identify and select mesh elements with minimal separation in the observer’s projected view plane.

Light-travel-time effects, which can affect binary-system analysis, are not yet included in the current implementation \citep[e.g.,][]{ltte_hw_virginis, ltte_hu_aquarii, ltte_moa}. This remains a key limitation and will be addressed in future releases. Other close-binary effects—such as mutual reflection and tidal distortions—are likewise absent from the modelling workflow.

Integrating \spice\ with \phoebe\ helps overcome these limitations. Mesh geometries, visibility maps, and surface parameters (e.g., gravity and temperature) can be imported directly, enabling accurate modelling of detached binaries in which \phoebe\ already accounts for tidal effects and external irradiation. \phoebe\ also supplies meshes and visibility information that include self-occlusion, a capability not yet implemented in the current \spice\ occlusion-detection algorithm. Further details are provided in Section~\ref{sec:phoebe_integration}.

\subsection{Spectral synthesis}
\label{sec:spectrum_emulator}

\spice\ relies on an accurate model of the specific intensity emitted by each stellar surface element, as expressed in Equation~\ref{eq:synthetic_flux}, which sums over the surface mesh. In principle, one could compute these intensities directly by running a full spectral synthesis for every mesh element with existing codes. In practice, this is computationally prohibitive due to (i) the large number of mesh triangles and (ii) the need to model high-resolution spectral time series. To overcome this, \spice\ uses emulators trained on grids of specific intensities, which amortise the cost of direct synthesis and are well-suited to both CPU and GPU execution, thereby improving efficiency.

\spice\ maintains flexibility with respect to the synthetic‑spectrum model employed, because the mechanism for calculating specific intensities is not embedded within the code. This design allows users to choose among existing spectral‑synthesis tools capable of intensity calculations, provided these tools can be configured to output specific intensities parameterised by inclination angle, wavelength, and atmospheric parameters. Here, we use the recently developed transformer‑based model \texttt{TransformerPayne} \citep{tpayne2024}. \spice\ also supports interpolation on traditional grids of specific intensities when such grids are supplied together with a suitable interface.

We emphasise that, unlike traditional stellar-spectroscopy applications that utilise flux measurements, \spice\ requires specific intensities at discrete viewing angles, not angle-averaged flux values. This requirement means users cannot simply employ existing grids of stellar fluxes, which represent radiation integrated over a homogeneous stellar surface. Instead, the emulator must predict specific intensities, radiation in a particular direction, because these retain the angular dependence essential for modelling stellar surfaces that are not homogeneous in their velocity fields and parameters (effective temperature, surface gravity, etc.).

\begin{table*}
    \caption{Photometric systems implemented in SPICE and adopted system response functions.}
    \label{tab:passbands}
    \begin{tabular}{lcc}
        \hline
        \textbf{System} & \textbf{Passbands} & \textbf{Transmission curves} \\
        \hline
        Johnson-Cousins & U, B, V, $\rm{R}_c$, $\rm{I}_c$ & \cite{BesselMurphy2012} \\
        Str\"omgren & u, v, b, y & \cite{Bessel2011} \\
        Gaia & G, BP, RP & \cite{GaiaEDR3Riello} \\
        Gaia & RVS & \cite{GaiaRVS} \\
        SDSS & u, g, r, i, z & \cite{SDSSDoi} \\
        2MASS & J, H, $\rm{K}_s$ & \cite{Cohen2MASS} \\
        GALEX & NUV, FUV & \cite{GALEX} \\
        LSST & u, g, r, i, z, y & \cite{lsst_throughputs} \\
        Pan-STARRS1 & g, i, r, w, y, z, open & \cite{PanSTARRS1_Tonry_2012} \\
        Hipparcos-Tycho & $\rm{H}_p$, $\rm{B}_T$, $\rm{V}_T$ & \cite{BesselMurphy2012} \\
        \hline
    \end{tabular}
\end{table*}

Synthetic spectra can be integrated over wavelengths using a given transmission curve to produce synthetic photometry. This is done by \spice\ for the photometric systems summarised in Table \ref{tab:passbands}, using the photo-counting and energy integration as appropriate for each filter following the formalism discussed in \cite{Casagrande_VandenBerg_2014}.

\subsection{Differentiable programming using JAX}

The software is developed entirely in the \jax\ framework \citep{jax2018github}, a state-of-the-art Python package primarily used for machine learning applications and scientific computing due to its computational efficiency and interface similarity to the \numpy\ API \citep{numpy}.
Beyond computational efficiency, which is demonstrated in Section \ref{sec:efficiency}, \jax\ offers full automatic differentiation. This capability creates opportunities for gradient-based and machine-learning-based methods of parameter inference from photometric and spectroscopic time series. Such experimental approaches are reserved for future work.
A significant advantage of the \jax\ framework is its compatibility with both GPU and CPU architectures. GPU implementation offers substantial computational speed-up, while CPU processing trades faster execution for accessibility, potentially requiring reduced-accuracy spectroscopic emulators to achieve reasonable computation times.

\section{Results}
\label{sec:results}

To illustrate the capabilities of \spice\ across a range of astrophysical scenarios, we present three representative case studies: a spotted, rotating star; a pulsating star; and an eclipsing binary system. Each example demonstrates the framework’s ability to model complex surface inhomogeneities, temporal variability, and observational signatures in both photometric and spectroscopic domains. The first case focuses on a chemically peculiar star with localised abundance spots, building on top of a test case from the literature to validate the treatment of surface features and rotational modulation in synthetic spectra.

\subsection{Spectral lines in spotted star}
\label{sec:spotted_star}
\begin{figure}
    \centering
    \includegraphics[width=\columnwidth]{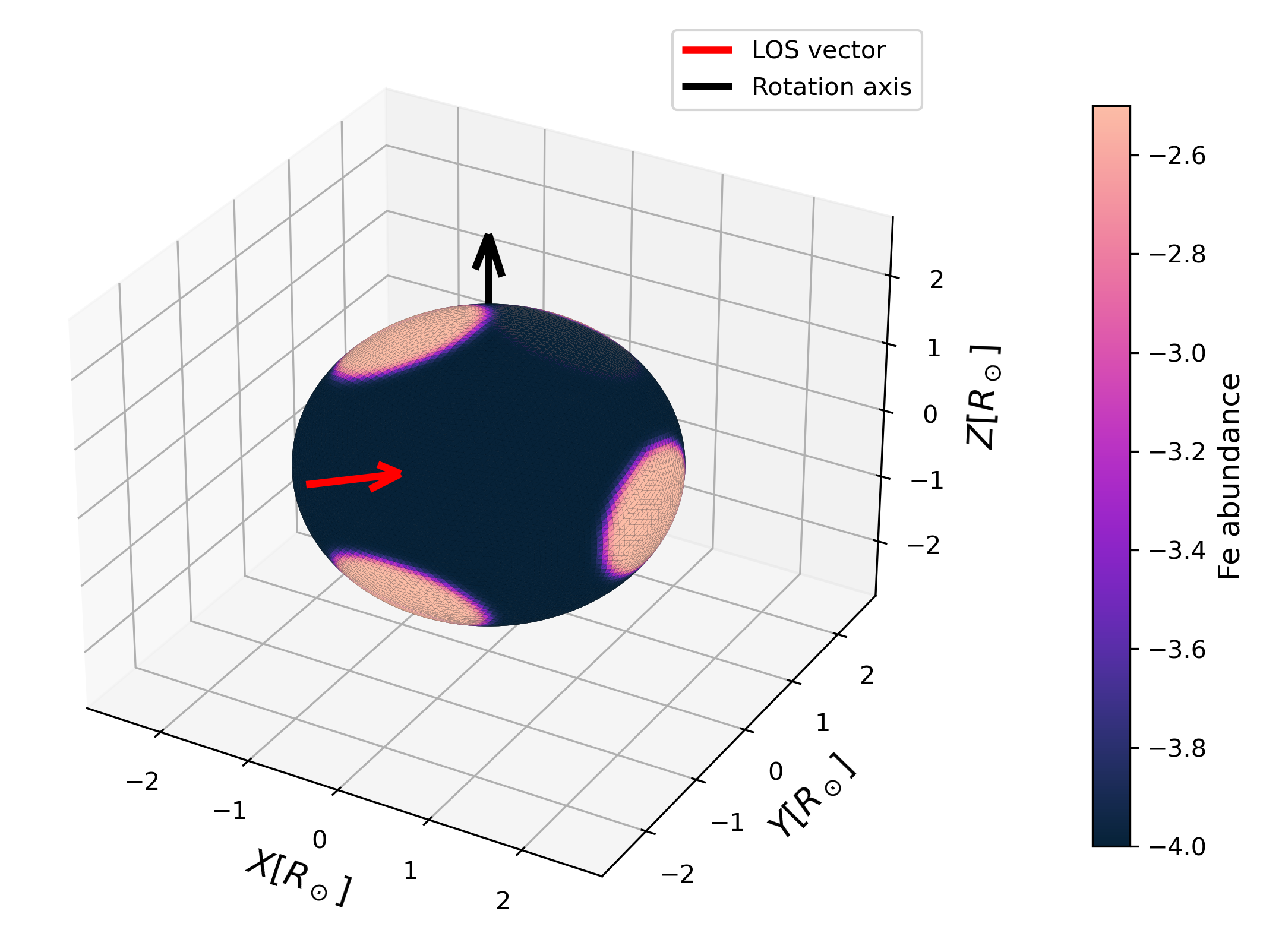}
    \caption{Mesh model reproducing one of the Doppler-imaging test cases presented by \citet{Kochukhov_2017_Doppler_Imaging_ApBp}. The model is shown in the star’s reference frame, with the line-of-sight (LOS) vector indicating the inclination under which the star is observed.}
    \label{fig:fe_spot_kochukov}
\end{figure}

\begin{figure}
    \centering
    \includegraphics[width=\columnwidth]{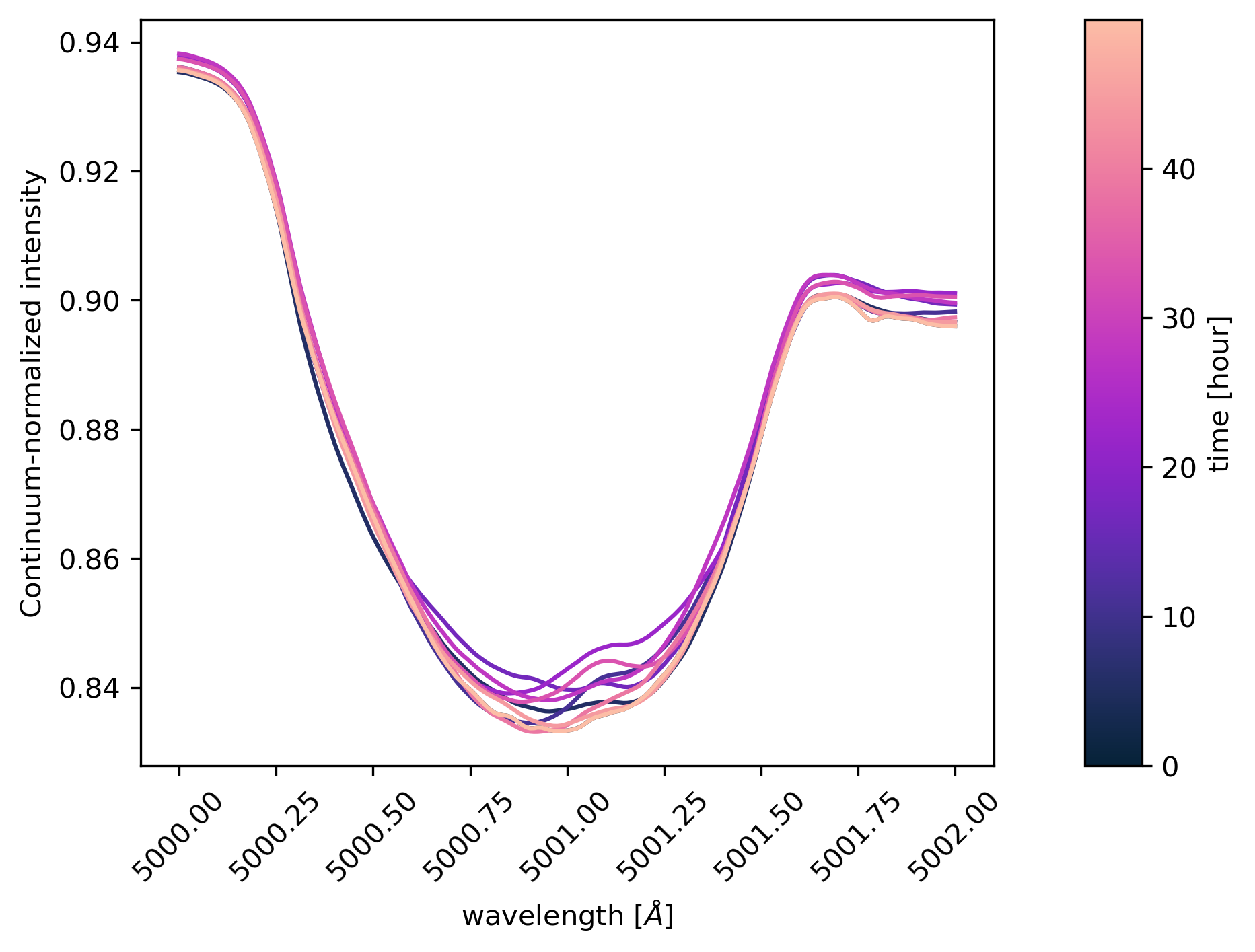}
    \caption{Synthetic Fe\,\textsc{i} line-profile variations over rotational phase for the spotted-star configuration described in Section~\ref{sec:spotted_star}. The star is modelled with a 20,480-element mesh and an inhomogeneous iron-abundance distribution following \citet{Kochukhov_2017_Doppler_Imaging_ApBp}. The spectrum has been post-processed with a 1D Gaussian kernel for clarity.}
    \label{fig:fe_line_kochukov}
\end{figure}

An example of a spotted, rotating star influencing the spectral lines was done by executing a 20,480-element mesh resembling the example presented by \cite{Kochukhov_2017_Doppler_Imaging_ApBp}. The model has a solar radius, $\log g$ = 4.0, $T_{\text{eff}}=8000$ K, an inclination of $60^\circ$ and a linear rotation velocity measure of $v \sin i $ of 40 km/s. The abundances are assumed to be solar values, which are adapted from \citep{Asplund_2021}, except of Iron, which is set to $[Fe/H]=-4.0$, with significant Iron overabundance spots of $\Delta[Fe/H]=-2.5$ set at latitudes of $-30^\circ$, $0^\circ$, $30^\circ$ and $60^\circ$, and at equidistant spans of longitudes. The mesh model is shown in Figure \ref{fig:fe_spot_kochukov}. We chose one of the Iron lines studied in the paper to reproduce and uniformly sampled 200 wavelengths between $\lambda_s=5000\ \mathring{A}$ and $\lambda_f=5002\ \mathring{A}$ (corresponding to a wavelength spacing of $\Delta \lambda \approx 0.01\ \mathring{A}$, i.e. resolution of $R \sim 5 \times 10^{5}$ at $\lambda \sim 5000\ \mathring{A}$), spanning 100 timestamps between 0 and 60 hours, covering one full rotation period. The resulting changing line profile is shown in Figure \ref{fig:fe_line_kochukov}. The spectrum in Figure \ref{fig:fe_line_kochukov} has been post-processed with a 1D Gaussian kernel smoothing for clarity.

\subsection{Pulsating star}
\label{sec:pulsating_star}
\begin{figure}
    \centering
    \includegraphics[width=\columnwidth]{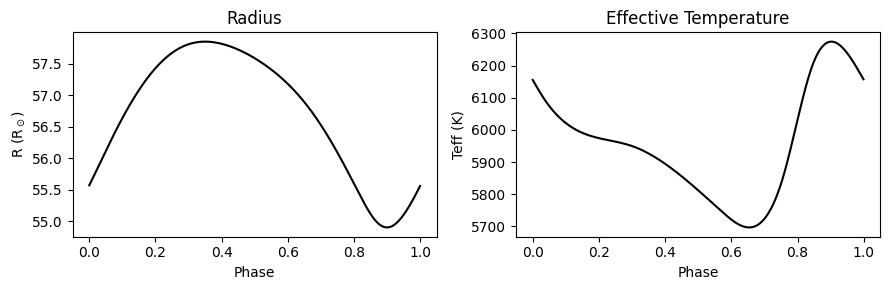}
    \caption{Radius and effective-temperature variations as a function of pulsation phase for the Cepheid model described in Section~\ref{sec:pulsating_star}, computed with \spips. These profiles are used as inputs to the \spice\ model to simulate the corresponding photometric and spectroscopic variability.}
    \label{fig:v_cygni_template}
\end{figure}

To demonstrate the use of \spice\ on pulsating stars, we focus on a Cepheid use case. These stars are particularly advantageous because their radial pulsations can be effectively modelled using a single mode, which is sufficient to produce realistic results.
The relationship between radius, temperature, and log g was computed for the parameters of $\delta$ Cephei using the \spips\ code \citep{spips} and decomposed into a Fourier-series approximation. The resulting radius and temperature profiles shown in Figure \ref{fig:v_cygni_template} were used to simulate photometric and spectroscopic variations with \spice. A 5,120-element model was used for this test.

\begin{figure}
    \centering
    \includegraphics[width=\columnwidth]{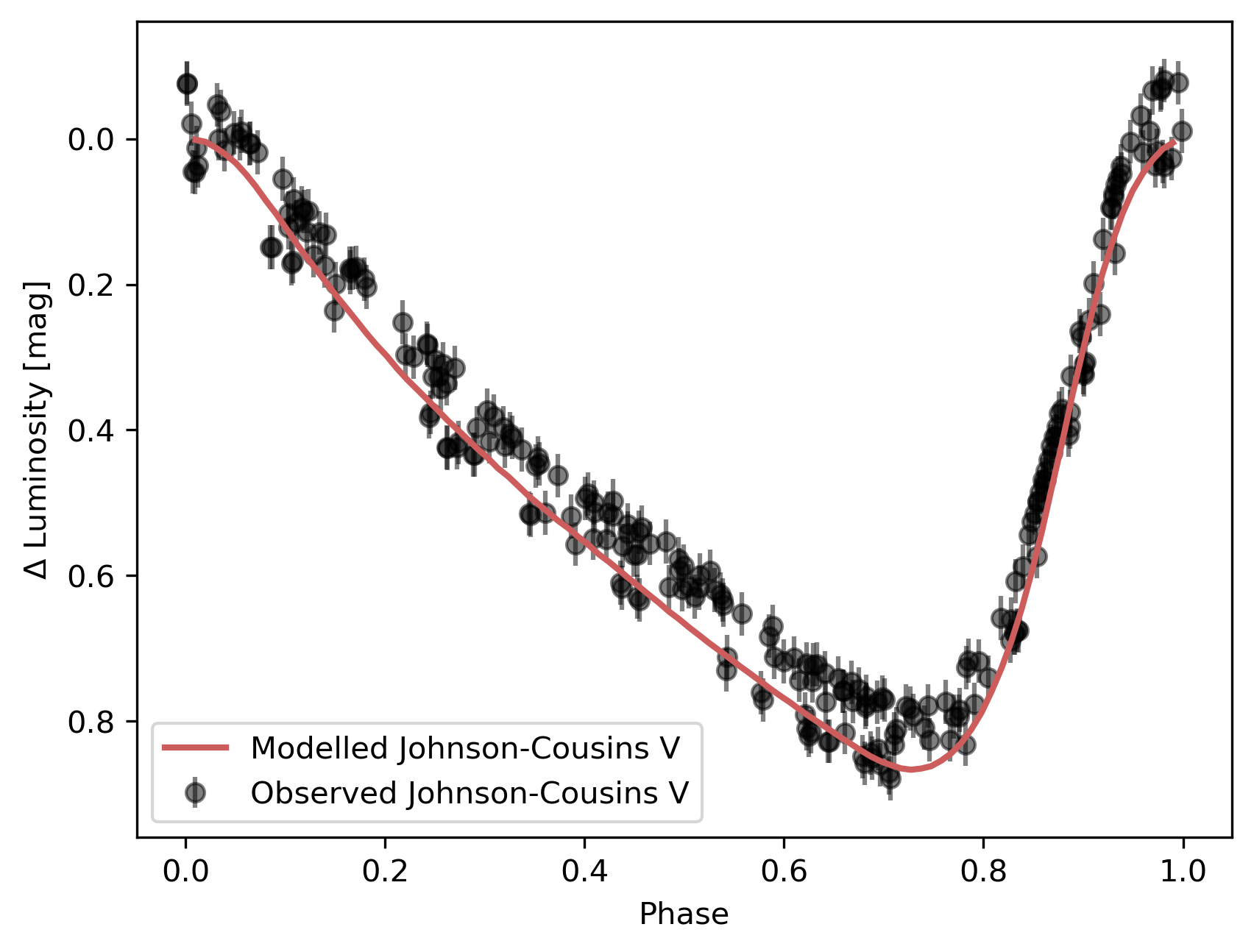}
    \caption{Synthetic Johnson--Cousins $V$-band light curve generated with \spice\ compared with the observational $V$-band data for $\delta$~Cephei used in \spips\ for the Cepheid model of Section~\ref{sec:pulsating_star}. The synthetic curve is computed using the phase-dependent radius and temperature profiles shown in Figure~\ref{fig:v_cygni_template}, and the observations are compiled from \citet{Moffett1984}, \citet{Kiss1998}, \citet{Berdnikov2008}, and \citet{Engle2014}.}
    \label{fig:cepheid_lightcurve}
\end{figure}

Figure \ref{fig:cepheid_lightcurve} presents a comparison between a synthetic light curve generated by \spice\ in Johnson-Cousins V band and observational data for $\delta$ Cephei, using the temperature and radius changes provided by \spips\ example. The observational data are compiled from \cite{Moffett1984, Kiss1998, Berdnikov2008, Engle2014}. 

\begin{figure}
    \centering
    \includegraphics[width=\columnwidth]{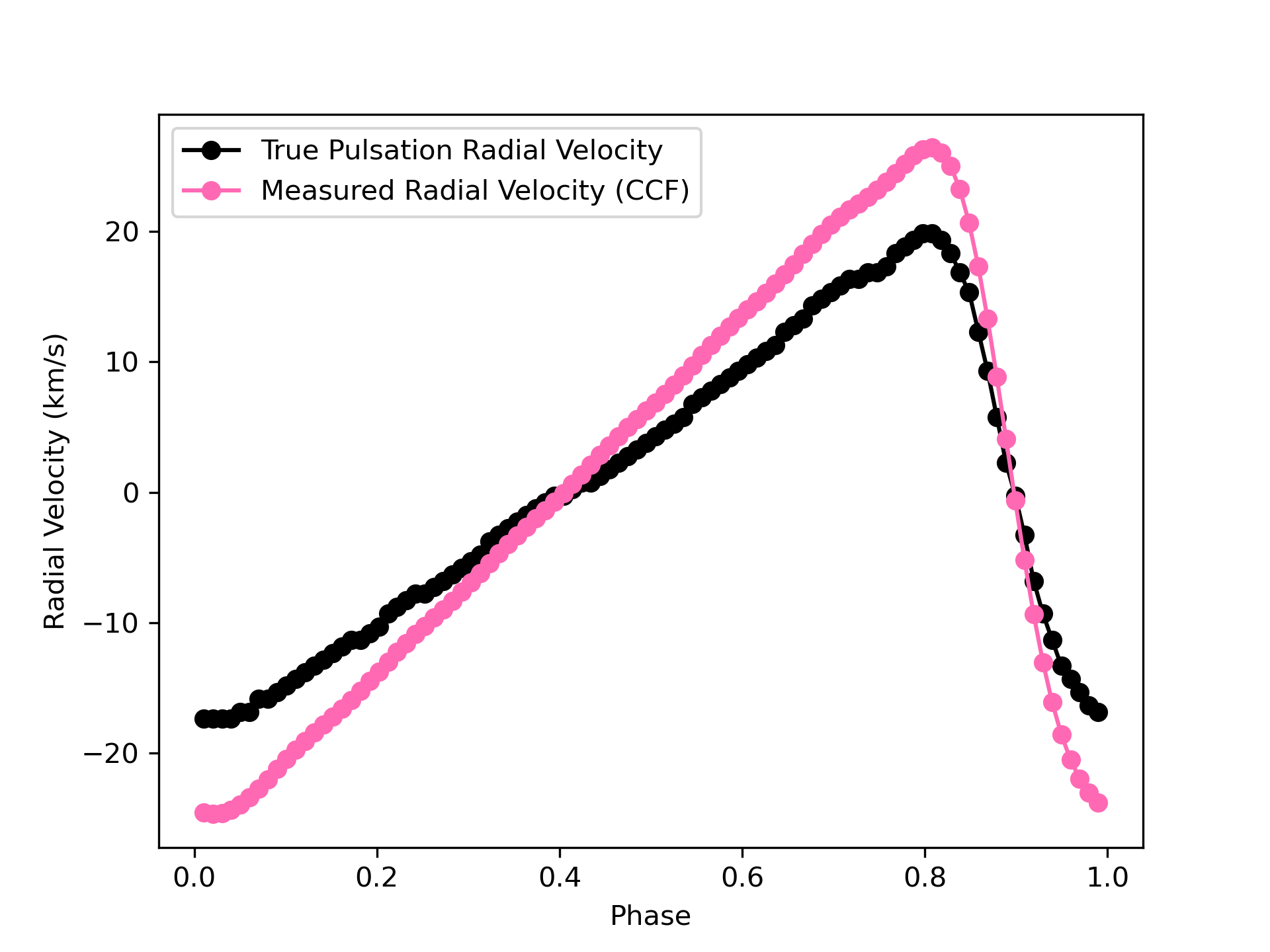}
    \caption{Comparison between radial velocities inferred from a cross-correlation function (CCF) applied to synthetic spectra and the true pulsation velocities of the mesh for the Fe\,\textsc{i} 4924.77~\AA\ line in the Cepheid model of Section~\ref{sec:pulsating_star}. The ratio of the true pulsation velocity to the CCF-based radial velocity has a mean value of 1.48, close to the expected factor of 1.5 from geometric projection arguments.}
    \label{fig:radvel_ccf}
\end{figure}

A validation of geometric properties of the pulsations was done by simulating a time series of spectra around a full pulsation cycle focused on a selected Fe I line (for our example, it was a sample of intensities computed for an equally sampled array of 2000 wavelengths in a 4 $\mathring{A}$-window around the Fe I line at 4924.77 $\mathring{A}$), calculating a synthetic radial velocity observation time series using the cross-correlation function, and comparing that to the true pulsation radial velocity which can be calculated from the coordinates of the mesh model for consecutive time steps. This quantity is often referred to as the p-factor, which plays a crucial role in converting observed spectroscopic radial velocities into the true photospheric pulsation velocities. Accurate determination of the p-factor is essential for methods such as the Baade–Wesselink technique, as it directly influences the inferred distances to Cepheid variables and thus the calibration of the cosmic distance scale \citep{di2013towards, nardetto2017harps}. The resulting comparison is shown in Figure \ref{fig:radvel_ccf}. The mean of the true pulsation velocity and mock observed radial velocity ratio is 1.48, which is close to the expected 1.5 ratio resulting from the geometric projection. Different spectral lines resulted in different p-factor values, which mostly aligned with usual p-factor measurements of cepheid stars \citep{merand_delta_cephei}. The results obtained using the synthetic spectra centred around specific lines are shown in Table \ref{tab:p_factor_values}. We saw negligible differences between p-factor values calculated using spectra obtained for models with different mesh resolutions. We note that the p-factor values differ between various lines, and it has been previously demonstrated that the resulting p-factor estimate may depend on the line choice for cross-correlation function calculation \citep{nardetto2013understanding}.

\begin{table}
    \centering
    \caption{p-Factor values obtained by comparing the mesh model pulsation velocity and results obtained by calculating the cross-correlation function using synthetic spectra for different lines.}
    \begin{tabular}{lc}
        \hline
        \textbf{Line Center} & \textbf{p-factor} \\
        \hline
        Fe I 4896.44 $\mathring{A}$ & 1.29 \\
        Fe I 4924.77 $\mathring{A}$ & 1.48 \\
        Fe I 5044.21 $\mathring{A}$ & 1.36 \\
        Fe I 5049.82 $\mathring{A}$ & 1.30 \\
        Si II 6347.00 $\mathring{A}$ & 1.30 \\
        Si II 6371.00 $\mathring{A}$ & 1.63 \\
        \hline
    \end{tabular}
    \label{tab:p_factor_values}
\end{table}

\subsection{Binary system}
\label{sec:binary_system}

\begin{figure}
    \centering
    \includegraphics[width=\columnwidth]{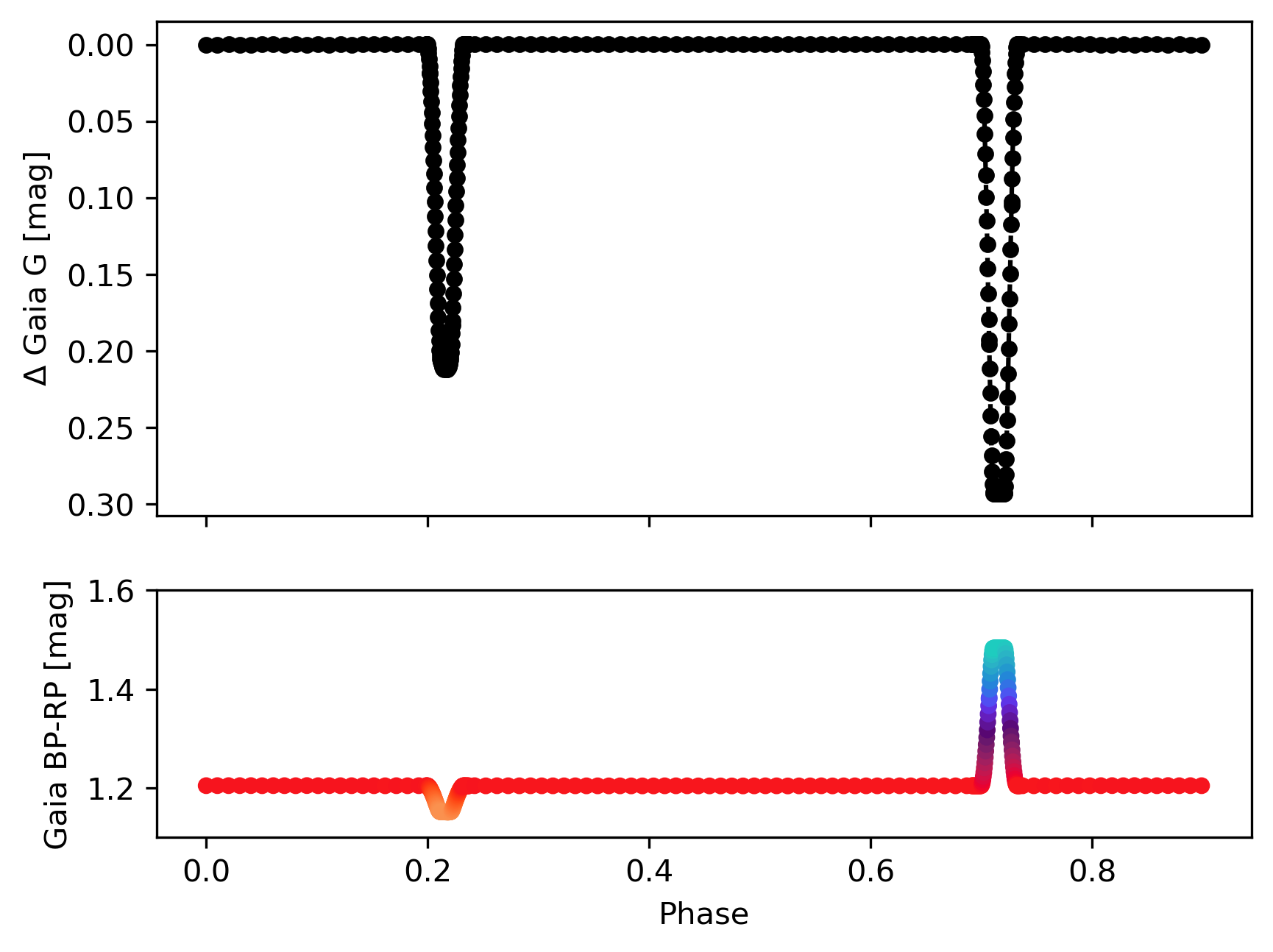}
    \caption{Synthetic Gaia $G$-band light curve computed from \spice\ synthetic spectra for the TZ Fornacis binary model described in Section~\ref{sec:binary_system}. The light curve is calculated over 300 time steps spanning one full orbital period.}
    \label{fig:tz_fornacis_lc}
\end{figure}

\begin{figure*}
    \centering
    \includegraphics[width=\linewidth]{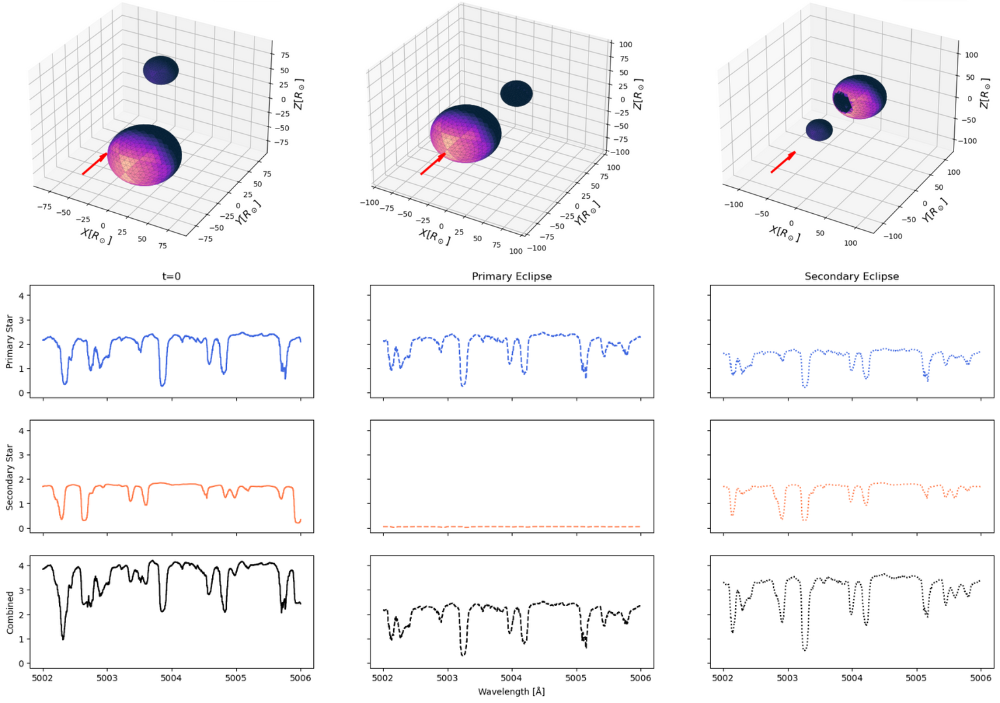}
    \caption{Synthetic spectra for the TZ~Fornacis model of Section~\ref{sec:binary_system} at $t = 0$ and during both eclipses. Colours on the mesh indicate the projected visible surface area of each element. The subtle patterns on the surface arise from small variations in mesh-triangle areas introduced by successive icosphere subdivisions. In the primary (deeper) eclipse, the secondary’s flux contribution vanishes, whereas in the secondary (shallower) eclipse, the primary is only partially occulted.}
    \label{fig:tz_fornacis_spectra}
\end{figure*}

\begin{figure}
    \centering
    \includegraphics[width=\columnwidth]{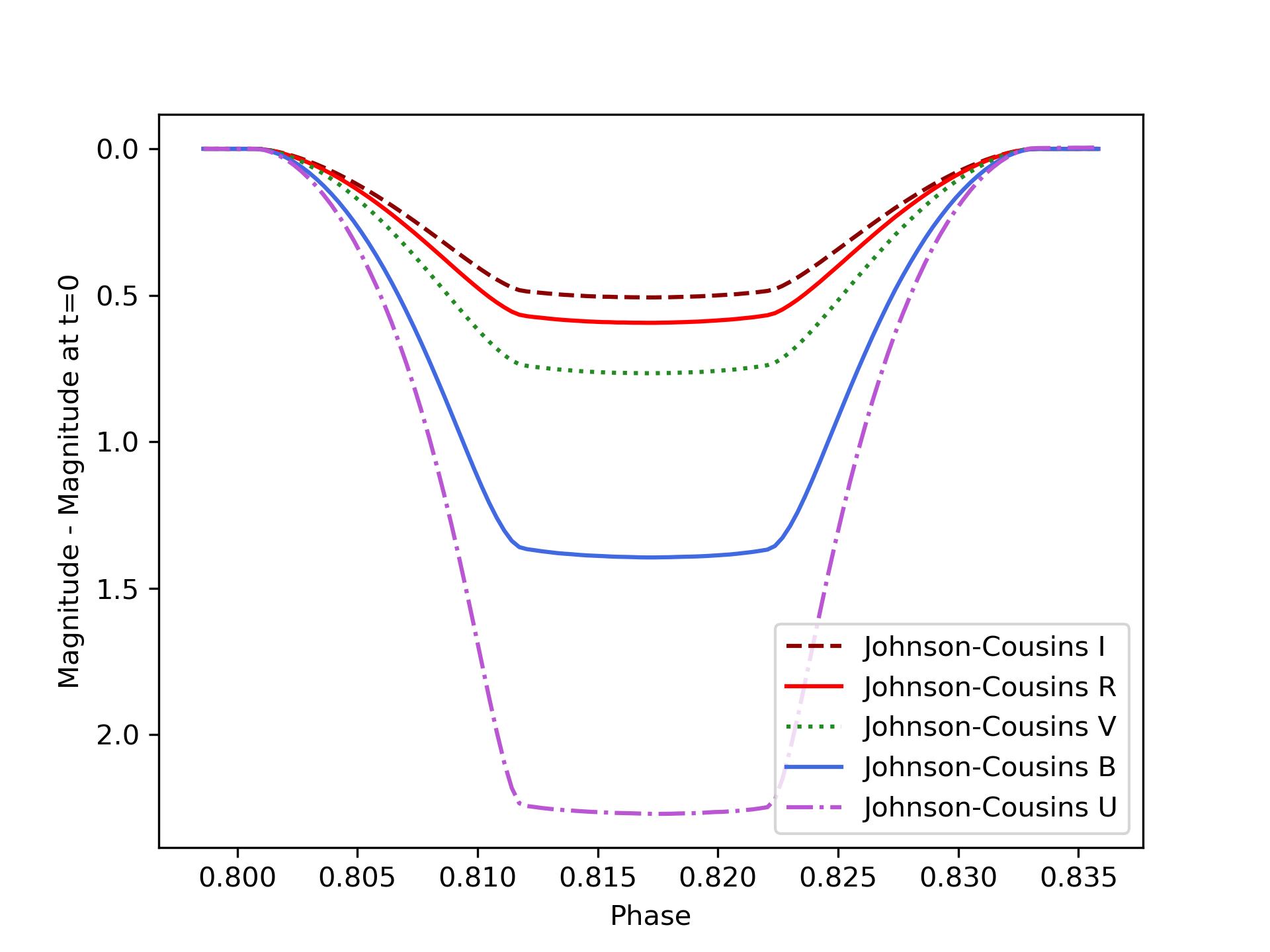}
    \caption{Multi-band light curve of the secondary eclipse for the TZ~Fornacis model discussed in Section~\ref{sec:binary_system}. The eclipse profile becomes steeper in bluer passbands and progressively flatter toward the infrared, reflecting the wavelength dependence of limb darkening.}
    \label{fig:tz_fornacis_eclipse}
\end{figure}

To fit within the range of parameters currently supported by \tpayne, we have selected a wide, fully detached eclipsing binary system with existing observations, consisting of both components with parameters within the grid. The system we have chosen for the simulation is TZ Fornacis, with the orbital values and stellar parameters adapted from \cite{tz_fornacis_orbit}. The parameter values are shown in Table \ref{tab:tz_fornacis_parameter_values}. The mesh models and orbit were fully modelled with \spice, with no \phoebe\ integration. The synthetic light curve was calculated for 300 timesteps spanning one orbit period. The Gaia G-band synthetic light curve is shown in Figure \ref{fig:tz_fornacis_lc}, while Figure \ref{fig:tz_fornacis_spectra} presents simulated spectra at $t=0$ and during both eclipses. The spectra during eclipses show expected behaviour: in the primary (deeper) eclipse, the secondary star’s contribution drops to zero, whereas in the secondary (shallower) eclipse, the primary’s flux decreases only partially, reflecting the partial occultation of its surface. A zoomed-in view of the secondary eclipse light curve is shown in Figure \ref{fig:tz_fornacis_eclipse}. With properly implemented limb darkening and partial occultation of the primary star, the eclipse profile is expected to be steeper in bluer filters, while becoming progressively flatter toward the infrared due to the reduced strength of limb darkening at longer wavelengths. This wavelength dependence is more commonly highlighted in the context of planetary transits, and its implications for \spice\ applications in exoplanet detection will be explored in future work.

\begin{table}
    \centering
    \caption{Parameters of the TZ Fornacis model used in the test case}
    \begin{tabular}{lcc}
        \hline
        \textbf{Parameter} & \textbf{Value} \\
        \hline
        Primary Radius & 8.28 $R_\odot$ \\
        Primary Mass & 2.06 $M_\odot$ \\
        Primary $T_{eff}$ & 4930 K \\
        Primary $log\ g$ & 2.91 \\
        Secondary Radius & 3.94 $R_\odot$ \\
        Secondary Mass & 1.96 $M_\odot$ \\
        Secondary $T_{eff}$ & 6650 K \\
        Secondary $log\ g$ & 3.35 \\
        Period & $75.6$ days \\
        Argument of Periastron & $85.68^\circ$ \\
        Longitude of the Ascending Node & $65.99^\circ$ \\
        Mean Anomaly & $269.00^\circ$\\
        \hline
    \end{tabular}
    \label{tab:tz_fornacis_parameter_values}
\end{table}

\subsection{PHOEBE integration}
\label{sec:phoebe_integration}

\spice\ is integrated with the \phoebe\ package, which allows the user to import a mesh model generated by \phoebe, which uses an approach similar to ours while keeping the mesh more adaptive than an icosphere (approach adopted in the current version of \spice), and calculate the synthetic flux and photometry using our models. \phoebe\ is a software package designed for modelling and analysing binary and single-star systems, with a primary focus on photometric data. The software models numerous physical phenomena relevant to stellar systems, including heating and reflection effects in close binaries, mass transfer in semi-detached systems, stellar spots, and tidal distortions arising from gravitational interactions between binary components \citep{phoebe}. However, \phoebe\ is not designed for synthetic spectra calculation, pulsation modelling, and the surface inhomogeneities are limited to temperature and generalised metallicity (with no option of adding singular abundance spots). By combining \spice\ with \phoebe, we aim to leverage the numerous functionalities existing within \phoebe\ and build upon the possible use cases.

We compared synthetic photometric light curves generated by \spice\ and \phoebe\ using a blackbody spectrum in both cases. This choice ensures a strict numerical comparison and avoids discrepancies arising from the distinct flux grids used by the two codes. In the \phoebe\ model, limb darkening was manually disabled, since \spice\ does not apply limb darkening to blackbody fluxes (this effect is included when using a spectrum emulator). In contrast, \phoebe\  adopts an analytic limb-darkening law; both light-travel-time and irradiation effects were also disabled for consistency. Further details of the \phoebe\ configuration are provided in Appendix~\ref{sec:appendix_phoebe_integration}.

We reconstructed the 3D mesh models using the \spice\ implementation, which may produce slightly different mesh element counts. Synthetic fluxes were then computed for all \phoebe\ mesh elements using the same parameter set as in the corresponding icosphere model constructed in \spice. Finally, synthetic light curves were calculated across a grid of system parameters to compare the two codes under varied conditions, with orbital periods between $P = 0.25$ and $P = 10$ days, inclinations from $80^\circ$ to $90^\circ$, mass ratios from 0.5 to 1.0, and primary masses between $1\,M_\odot$ and $7\,M_\odot$.

\begin{table}
    \centering
    \caption{Statistics of differential-magnitude residuals between light curves computed with \phoebe\ and \spice. The quantity $N$ denotes the number of elements in the \spice\ mesh model.}
    \begin{tabular}{ccccc}
        \hline
        \textbf{N} &\textbf{Mean} & \textbf{Median} & \textbf{Standard deviation} & \textbf{Max} \\
        \hline
        1280 & $5.44\cdot 10^{-3}$ & $4.94\cdot 10^{-3}$ & $4.08\cdot 10^{-3}$ & $1.82\cdot 10^{-2}$ \\
        5120 & $1.04\cdot 10^{-3}$ & $9.04\cdot 10^{-4}$ & $8.66\cdot 10^{-4}$ & $7.92\cdot 10^{-3}$ \\
        20480 & $3.93\cdot 10^{-4}$ & $2.99\cdot 10^{-4}$ & $4.19\cdot 10^{-4}$ & $3.86\cdot 10^{-3}$ \\
        \hline
    \end{tabular}
    \label{tab:filter_lightcurve_residuals}
\end{table}

We calculated synthetic filter magnitudes at 50 evenly spaced timestamps covering both the primary and secondary eclipses\footnote{The eclipse times were estimated using an algorithm that interpolates the relative sky-plane motion of two bodies with cubic polynomials and identifies eclipse contact points. The algorithm implementation is available in the code repository.} in the top-hat filter with unit transmission between 900~$\mathring{A}$ and 40,000~$\mathring{A}$ (\texttt{Bolometric900-40000} in \phoebe).  The residuals between \spice\ and \phoebe\ light curves were at most $10^{-3}$~mag, with mean differences on the order of $10^{-4}$~mag or smaller across all tested parameters for the mesh models with higher resolution. This level of offset is below typical observational uncertainties, demonstrating that \texttt{SPICE} can be reliably applied to binary systems in its current implementation. Table~\ref{tab:filter_lightcurve_residuals} summarises the residual statistics for each filter across varying mesh resolutions. The most significant residuals consistently occurred in systems with low mesh resolution (around 1000 mesh elements), likely arising from the different mesh construction schemes—and hence differing element sizes and counts—used in \spice\ and \phoebe.

\subsection{Efficiency}
\label{sec:efficiency}

The most computationally intensive step in spectrum generation is spectrum emulation. In our case, we are using \tpayne, and the method invoking the intensity emulation is the one that takes the most time in the whole procedure, which can be seen when comparing the time measurements with the simple Gaussian line model case, which is modelled using one analytical equation, thus introducing negligible computational overhead.

The test we ran consisted of calculating the fluxes at 1000 uniformly spaced wavelength values using \tpayne\ for mesh representations with varying numbers of elements. The time of computation does not depend on the choice of wavelengths, but just on the number of wavelengths for which the synthetic flux is computed; therefore, an arbitrary wavelength range was chosen. The flux calculation in \spice\ is performed using a chunking approach inspired by \texttt{FlashAttention}, which significantly reduces memory usage \citep{flashattention}. The chunking approach is explained in more detail in Appendix \ref{sec:appendix_code_optimizations}. Our results were calculated using the default batch size of 1024 wavelength values, a setting that offsets computation time and GPU memory requirements. The results, which are mean averages of five runs for each parameter configuration, are summarised in Table \ref{tab:execution-times}. The computation time scales linearly with the size of the wavelength array.

\begin{table*}
\centering
\caption{Benchmark timing comparison for the Tesla V100-SXM2-32GB (V100 in the Table) and A100-PCIE-40GB (A100 in the Table) GPUs, for \tpayne\ and a simple Gaussian line emulator. All times are in seconds.}
\begin{tabular}{lcc}
\hline
\textbf{Mesh resolution} & \textbf{A100} & \textbf{V100} \\
\hline
\multicolumn{3}{c}{\texttt{TransformerPayne}} \\
\hline
320   & $1.05 \pm 0.04$ & $2.97 \pm 0.01$\\
1280  & $3.08 \pm 2.07$ & $6.33 \pm 0.87$\\
5120  & $6.87 \pm 1.58$ & $18.28 \pm 0.86$\\
20480 & $21.95 \pm 1.24$ & $62.49 \pm 0.81$\\
\hline
\multicolumn{3}{c}{Gaussian Line Emulator} \\
\hline
320   &$0.038 \pm 0.004$ & $0.018 \pm 0.002$ \\
1280  &$0.33 \pm 0.56$ & $0.20 \pm 0.33$ \\
5120  &$0.34 \pm 0.52$ & $0.23 \pm 0.32$ \\
20480 &$0.53 \pm 0.57$ & $0.39 \pm 0.40$ \\
\hline
\end{tabular}
\label{tab:execution-times}
\end{table*}

\section{Conclusions}
\label{sec:conclusions}

We have presented \spice, an open-source software package for spectral synthesis of inhomogeneous stellar surfaces. \spice\ incorporates physical models to simulate a range of astrophysical scenarios: surface inhomogeneities (e.g., temperature and abundance spots), stellar pulsations, and eclipsing binary systems. We have validated the software on several test cases, demonstrating its capabilities and reliability in studying real case scenarios. The implementation provides a foundation for future observational studies and comparative analyses with empirical data. \spice\ is publicly available on GitHub and distributed via the Python Package Index (PyPI) for straightforward installation and deployment. The codebase is written entirely in Python and leverages the computational advantages of the \jax\ framework, including automatic differentiation and just-in-time compilation for optimised performance.

These features enable efficient gradient-based inference methods, positioning \spice\ as a scalable solution for processing large-scale spectroscopic surveys and analysing extensive datasets anticipated from current and upcoming astronomical instruments. The computational efficiency gained through these modern programming paradigms makes the tool particularly well-suited for the demanding requirements of contemporary stellar astrophysics research, where rapid analysis of high-volume photometric and spectroscopic observations is increasingly essential.

The modular architecture of \spice\ facilitates its extension to additional physical scenarios and enables seamless integration with \phoebe, thereby supporting the broader research community in stellar characterisation and population studies. 
With its capability to incorporate 
stellar inhomogeneities in parameter inference from spectra, \spice\ is well-positioned to harvest time-series information from spectroscopic surveys. 
Planned future developments include optimisations such as adaptive mesh construction, incorporation of light travel time effects (currently absent from the rigorous binary system treatment), Doppler boosting, and the addition of further stellar phenomena, including rotational flattening and differential rotation.
We anticipate \spice\ will be particularly useful for incorporating stellar inhomogeneities into parameter inference from spectra in extensive surveys, analysing pulsations from line profile variations, characterising rotational flattening from both photometric and spectroscopic time series, and many related applications. 

\section*{Acknowledgements}

We would like to kindly thank Amanda Barnard, Jackie Blaum, Josh Bloom, Sven Buder, Ben Montet, Benjamin Shappe, Ioana Ciuc\u{a}, Giacomo Cordoni, Kyle Conroy, Fei Dai, Dan Hey, Daniel Huber, Joel Ong Jia Mian, Katarzyna Kruszy\'{n}ska, Mark Krumholz, Rena Lee, Yaguang Li, Andrew McWilliam, Antoine Mérand, Melissa Ness, Cheng Soon Ong, Clauda Reyes, Nicholas Saunders, Hannah Schunker, Rachel Street, Ian Thompson, Jennifer van Saders, Stuart Wyithe, Jie Yu, and George Zhou for consultations, ideas for future development, and helpful comments.

\section*{Data Availability}

This study uses observational data available in the \spips\ code repository, with observations compiled from \citet{Moffett1984}, \citet{Kiss1998}, \citet{Berdnikov2008}, and \citet{Engle2014}, passband transmission curves from cited publications, and data simulated by \spice\ and \phoebe\ frameworks with codes available in the public \spice\ code repository.



\bibliographystyle{mnras}
\bibliography{example} 




\appendix

\section{Mesh Construction}
\label{sec:appendix_mesh_construction}

\begin{figure}
    \centering
    \includegraphics[width=\linewidth]{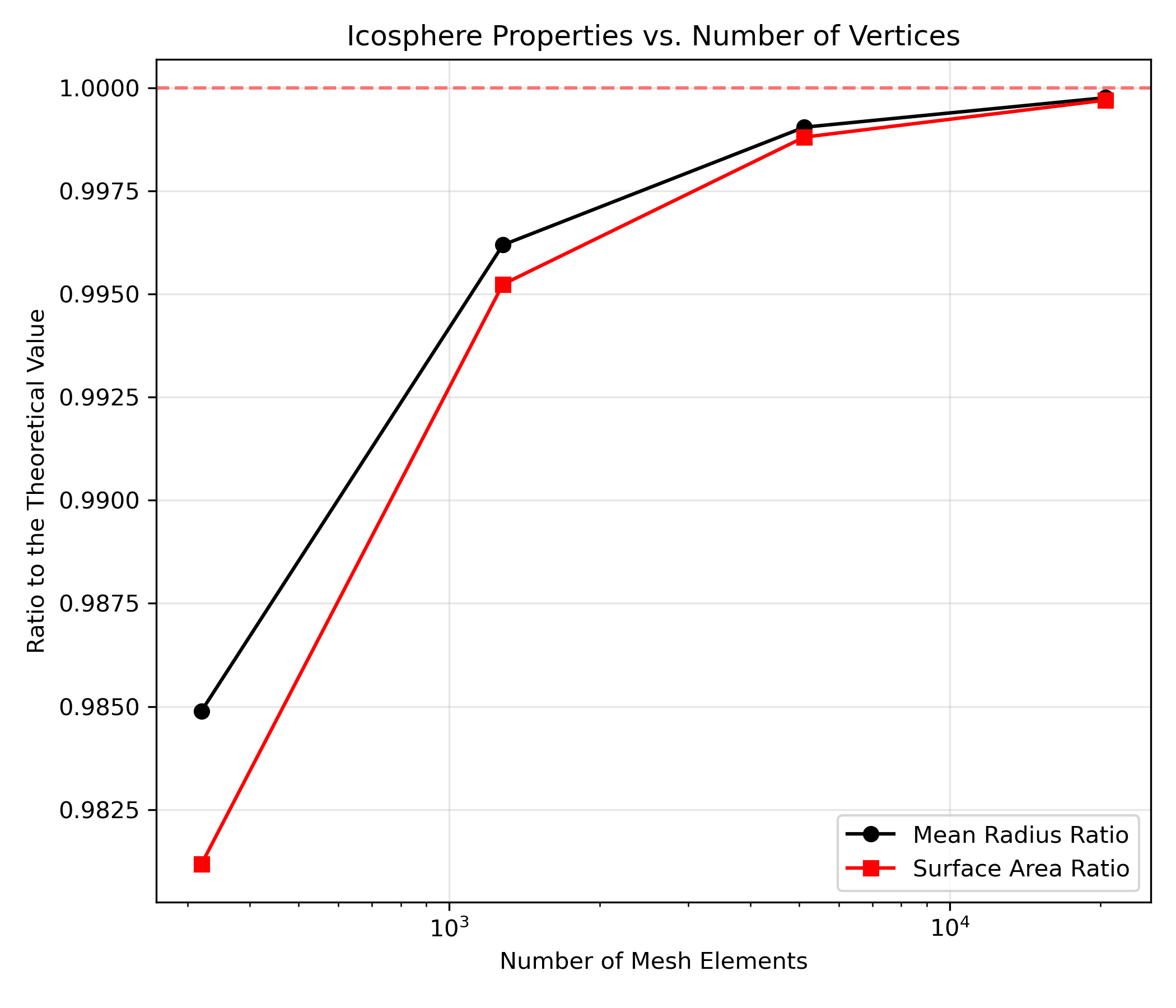}
    \caption{Ratios of the mean mesh-element radius to the modelled stellar radius $R$ (top) and of the total mesh surface area to the theoretical value $4\pi R^2$ (bottom), as a function of icosphere subdivision level. Both ratios converge to unity as the number of mesh elements increases, illustrating the accuracy of the icosphere approximation.}
    \label{fig:icosphere_value_ratios}
\end{figure}

As outlined in the main body of the paper, the 3D mesh model constructed in \texttt{SPICE} is an icosphere model. The number of vertices is restricted by the number of subdivisions performed, thereby limiting the possible mesh element counts to a few values, as detailed in Table \ref{tab:subdivs_mesh_triangles}. For each subdivision number, we compare the mean distance from the mesh centre for each mesh element to the modelled star's radius $R$, and the actual surface area of the mesh model to the expected surface area $4\pi R^2$. The results are presented in Figure \ref{fig:icosphere_value_ratios}. These values are independent of the stellar radius and approach unity as the number of mesh elements increases.

\subsection{Spot smoothness calculation}
\label{sec:appendix_spots_smoothness}

\begin{figure}
    \centering
    \includegraphics[width=\linewidth]{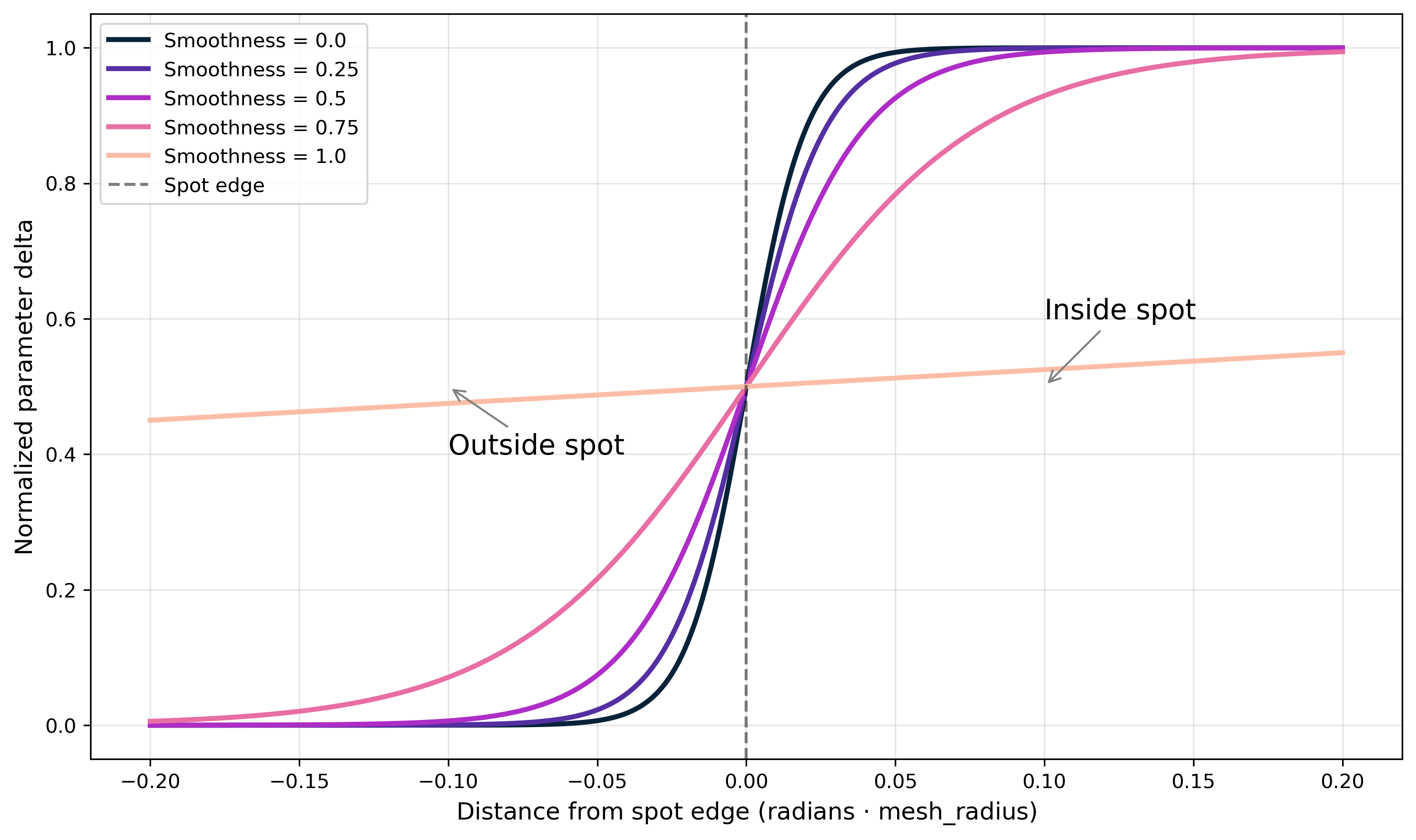}
    \caption{Normalised parameter value (expressed as the ratio between the spot value and the background value) as a function of distance from the spot centre for different smoothness parameters $\alpha$. Larger $\alpha$ values produce sharper spot boundaries, while smaller values yield more gradual transitions.}
    \label{fig:smoothness_function}
\end{figure}

\begin{figure}
    \centering
    \includegraphics[width=\linewidth]{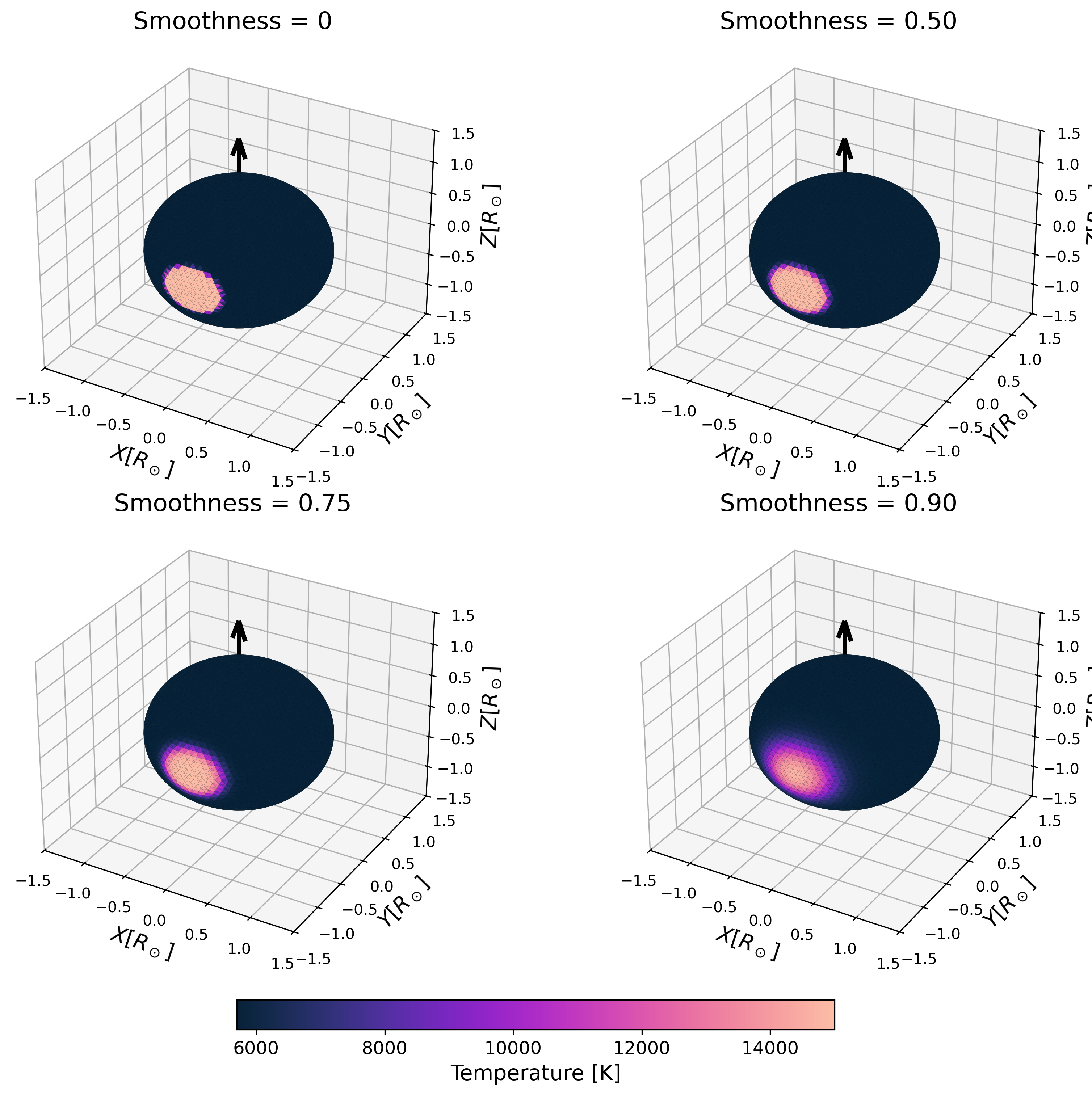}
    \caption{Examples of spots with different smoothness parameters $\alpha$ on a star with base temperature $T = 5700$~K and spot-centre temperature $T_\text{spot} = 15\,000$~K. Increasing $\alpha$ sharpens the spot edge, whereas lower values result in smoother, more diffuse boundaries.}
    \label{fig:spot_smoothness_examples}
\end{figure}

The smoothness parameter is rescaled to allow a more intuitive choice of the smoothness parameter $\alpha$ by the user. The rescaled form of the \texttt{sigmoid} function we use is of the following form:

\begin{equation}
    \Delta p_i = \Delta p_{\text{max}} \cdot \sigma\left(\frac{R_{\text{spot}} - d_i}{s \cdot R_{\text{mesh}}}\right),
\end{equation}

where $\sigma(x) = \frac{1}{1+e^{-x}}$ is the sigmoid function, $R_{\text{spot}}$ is the spot radius in degrees, $d_i$ is the cartesian distance from mesh point $i$ to the spot centre, $s \approx 0.01(1-\alpha) + 0.0001\alpha$ with $\alpha \in [0,1]$ being the smoothness parameter, and $R_{\text{mesh}}$ is the mesh radius. The effect of the smoothness parameter $\alpha$ on the spot boundary in terms of the parameter value is demonstrated in Figure \ref{fig:smoothness_function}, and the examples of spots with various smoothness parameter values are shown in Figure \ref{fig:spot_smoothness_examples}.

\section{Radiative quantities}
\label{sec:appendix_radiative_quantities}

\subsection{Bolometric luminosity}

\spice\ can calculate bolometric luminosity using a similar approach to the one outlined in Section \ref{sec:integrating_flux_from_mesh}, summing the flux contributions from all mesh elements (for bolometric luminosity, the summation occurs over all mesh elements $\forall_i A_i$, not only the visible ones):

\begin{equation}
L = \sum_{i=1}^{N_{\rm total}} \sum_{\lambda} F_\lambda(\vec{p}_i) \cdot A_i
\end{equation}

This calculation measures the energy output on the surface, and the result is independent of the observer's position, taking into account the output from all mesh elements. The key differences in the treatment of monochromatic luminosity and observed flux include:
\begin{itemize}
\item \textbf{Visibility consideration:} Flux calculations use only the visible portions of the stellar surface, whereas luminosity calculations integrate over the entire stellar surface.
\item \textbf{Directional dependence:} Flux calculations incorporate the projection effect through the viewing angle $\mu_i$, while luminosity calculations do not require this directional dependence.
\item \textbf{Distance scaling:} Observed flux includes an inverse-square distance dependence ($d^{-2}$), whereas luminosity is an intrinsic property independent of observer distance.
\end{itemize}

A computationally feasible method of calculating the flux from any given intensity function would be performing a sum:

\begin{equation}
F_\lambda(\vec p_i) \approx \sum_{m \in M} I_{\lambda}(\mathbf n_m)\,\mathbf n_m\,\Delta\Omega_m
\end{equation}

where $M$ is a set of selected values of $\mu$ spanning the range of $(-1, 1)$. Such an approximation, which involves integrating the intensity over a sufficiently large and evenly distributed set of directions, will be sufficient for many cases. However, for some simple cases, such as a blackbody spectrum, the flux can be expressed analytically as well.

\begin{equation}
\mathbf F_{\lambda}
      = \int_{4\pi} I_{\lambda}(\mathbf n)\,\mathbf n\,d\Omega
      \;\;\approx\;\;
      \sum_{m \in M} I_{\lambda}(\mathbf n_m)\,\mathbf n_m\,\Delta\Omega_m ,
\end{equation}

where $I_\lambda(\mathbf n)$ is the monochromatic specific intensity in direction $\mathbf n$, $\mathbf n$ is a unit vector on the unit sphere, $d\Omega$ is the differential solid-angle element, $M$ is the index set of discrete solid-angle elements, $\mathbf n_m$ is the unit vector pointing to the centre of element $m\in M$, and $\Delta\Omega_m$ is the solid angle of that element (with $\sum_{m\in M}\Delta\Omega_m \approx 4\pi$).

\subsection{Parallel-ray assumption}

\begin{figure}
    \includegraphics[width=\linewidth]{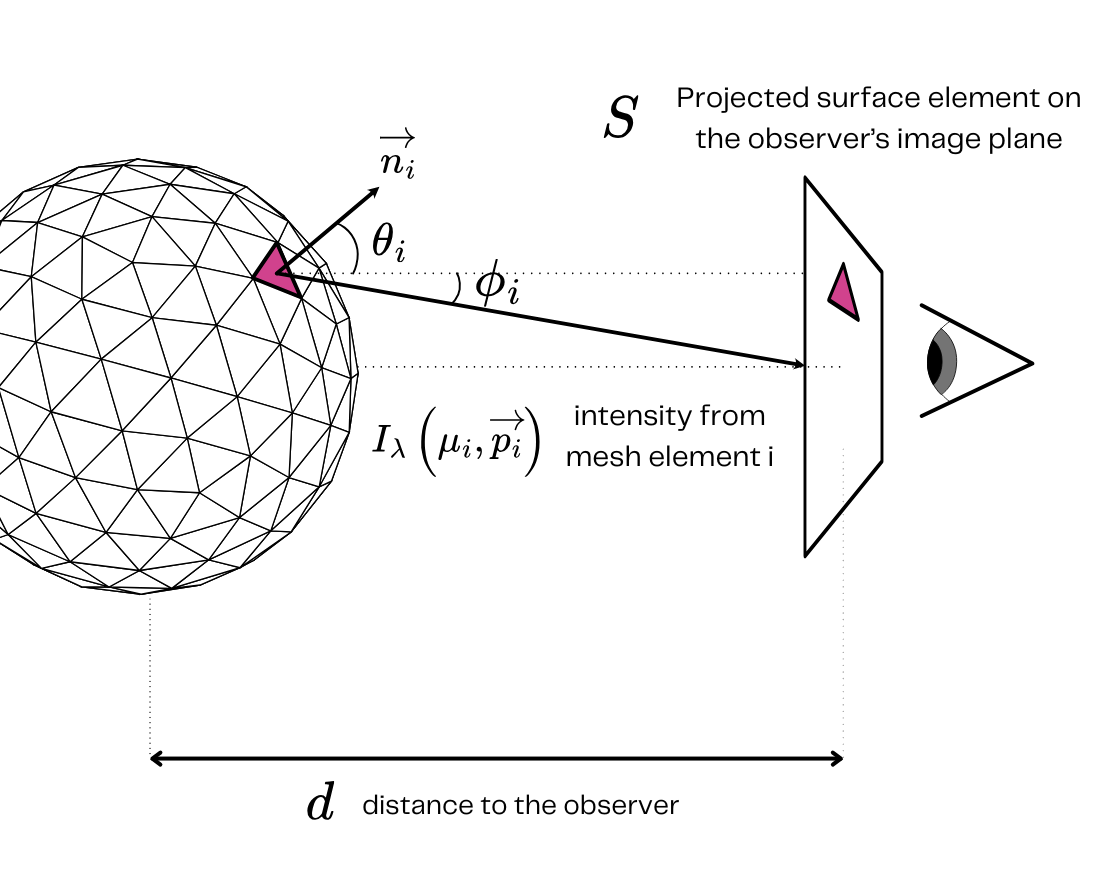}
    \caption{Synthetic observed-flux computation for a triangulated stellar surface without the parallel-ray assumption. In this more general geometry, the angle $\phi_i$ between the observer and the intensity ray modifies the effective viewing angle for off-centre mesh elements. \spice\ currently sets $\phi_i = 0$ for all elements, a simplification that has negligible impact for unresolved stellar discs.}
    \label{fig:appendix_spice_non_parallel_ray}
\end{figure}

Currently, we assume that the intensity rays are parallel to the observer. This simplification should not introduce any significant deviations to stars other than the Sun, as we usually can't resolve the stellar surfaces when observing them. However, it is worth noting we are neglecting the angle $\phi_i$ shown in Figure \ref{fig:appendix_spice_non_parallel_ray} and setting it to $\phi_i=0$ for all mesh elements. Even for the Sun, the angular size yields a cosine of the angular radius close to 1. Therefore, we consider this effect to be negligible.

\section{PHOEBE integration}
\label{sec:appendix_phoebe_integration}

\begin{table}
    \centering
    \caption{\texttt{PHOEBE} parameter values for the comparison system.}
    \begin{tabular}{lc}
        \hline
        \textbf{PHOEBE parameter} & \textbf{Parameter value} \\
        \hline
        \texttt{grav\_bol} & \texttt{0.} \\
        \texttt{ld\_mode} & \texttt{manual} \\
        \texttt{ld\_func} & \texttt{linear} \\
        \texttt{ld\_coeffs} & \texttt{[0.]} \\
        \texttt{ld\_mode\_bol} & \texttt{manual} \\
        \texttt{ld\_func\_bol} & \texttt{linear} \\
        \texttt{ld\_coeffs\_bol} & \texttt{[0.]} \\
        \texttt{atm} & \texttt{blackbody} \\
        \hline
    \end{tabular}
    \label{tab:binary_check_phoebe_parameters}
\end{table}

\texttt{PHOEBE} models several phenomena important for the correct treatment of close binary systems that are not yet covered by \spice. Some of the effects modelled by the 2.4.11 version of \texttt{PHOEBE} we used are: irradiation and reflection effects, gravity brightening, and light travel time effects \citep{PHOEBE_documentation}. 
Although some of these effects may be included in future \spice\ releases, the implementation of some of these effects is non-critical since the latter can import meshes directly from \phoebe.
Therefore, the \phoebe\ binary system configuration used to compare synthetic light curves is greatly simplified. The settings used for both components of the system are summarised in Table \ref{tab:binary_check_phoebe_parameters}. Computations are run without irradiation effects (\texttt{irrad\_method=none}) and without light travel time effects (\texttt{ltte=False}). Lightcurves are calculated using the \texttt{lc} (light curve) datasets and the following passbands: \texttt{Bolometric:900-40000}, \texttt{Gaia:G}, \texttt{Johnson:U}, \texttt{Johnson:V}, and \texttt{Str\"omgren:v}.

Since the \texttt{PHOEBE} version (2.4.11) we're using for these comparisons doesn't include Doppler shift effects in light-curve computations (a feature temporarily disabled in the used version due to concerns about accuracy), we manually disable them in \texttt{SPICE} for these experiments (using the \texttt{disable\_doppler\_shift} parameter in the \texttt{simulate\_observed\_flux} function for \texttt{PHOEBE}).

We compute synthetic passband magnitudes using both \texttt{PHOEBE} and \texttt{SPICE} with the following methods:

\begin{itemize}
    \item \texttt{PHOEBE}: We obtain time series of synthetic fluxes ($F_\mathrm{PHOEBE}$) by accessing the \texttt{fluxes@dataset@model} parameters (where \texttt{dataset} corresponds to the specific \texttt{lc} (light curve) dataset in a chosen passband). We calculate magnitudes for individual time steps as $m_\mathrm{PHOEBE,\ diff}(t) = -2.5\log_{10}\frac{F_\mathrm{PHOEBE}(t)}{F_\mathrm{PHOEBE}(t=0)}$, thus determining magnitude offsets relative to the synthetic flux at time $t=0$. This way, we can drop the dependency on zero-point values adopted by the flux calculation schemes in \texttt{PHOEBE} and \texttt{SPICE}.
    
    \item \texttt{SPICE}: We calculate time series of synthetic fluxes ($F_\mathrm{SPICE}$) using the \texttt{simulate\_observed\_flux} method for an array of 1000 wavelengths ranging from 900 \AA{} to 40000 \AA{}. We use these fluxes to calculate synthetic passband magnitudes with the \texttt{AB\_passband\_luminosity} function evaluated on the sum of synthetic fluxes calculated for both components separately. Similar to \texttt{PHOEBE}, differential magnitudes are calculated relative to the synthetic flux at time $t=0$.
\end{itemize}

\begin{figure}
    \centering
    \includegraphics[width=\linewidth]{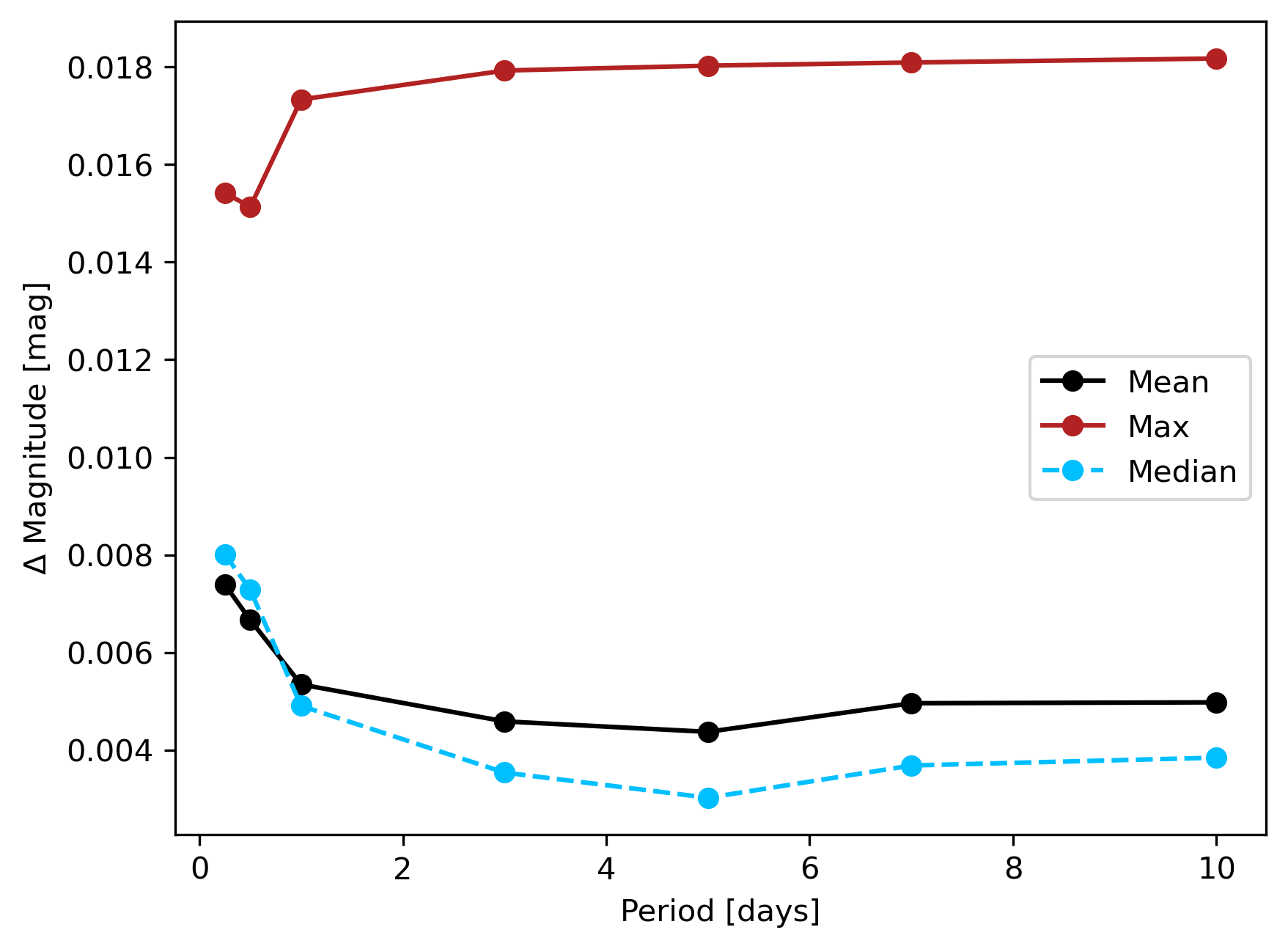}
    \caption{Maximum and minimum residuals in differential light curves $(t - t_0)$ between \phoebe\ and \spice\ for the low-resolution mesh case (1280 mesh elements in \spice, 1142 in \phoebe), as a function of orbital period. Residuals are aggregated over all explored parameter combinations (period, mass ratio, eccentricity, and primary mass).}
    \label{fig:appendix_residuals_1000}
\end{figure}

\begin{figure}
    \centering
    \includegraphics[width=\linewidth]{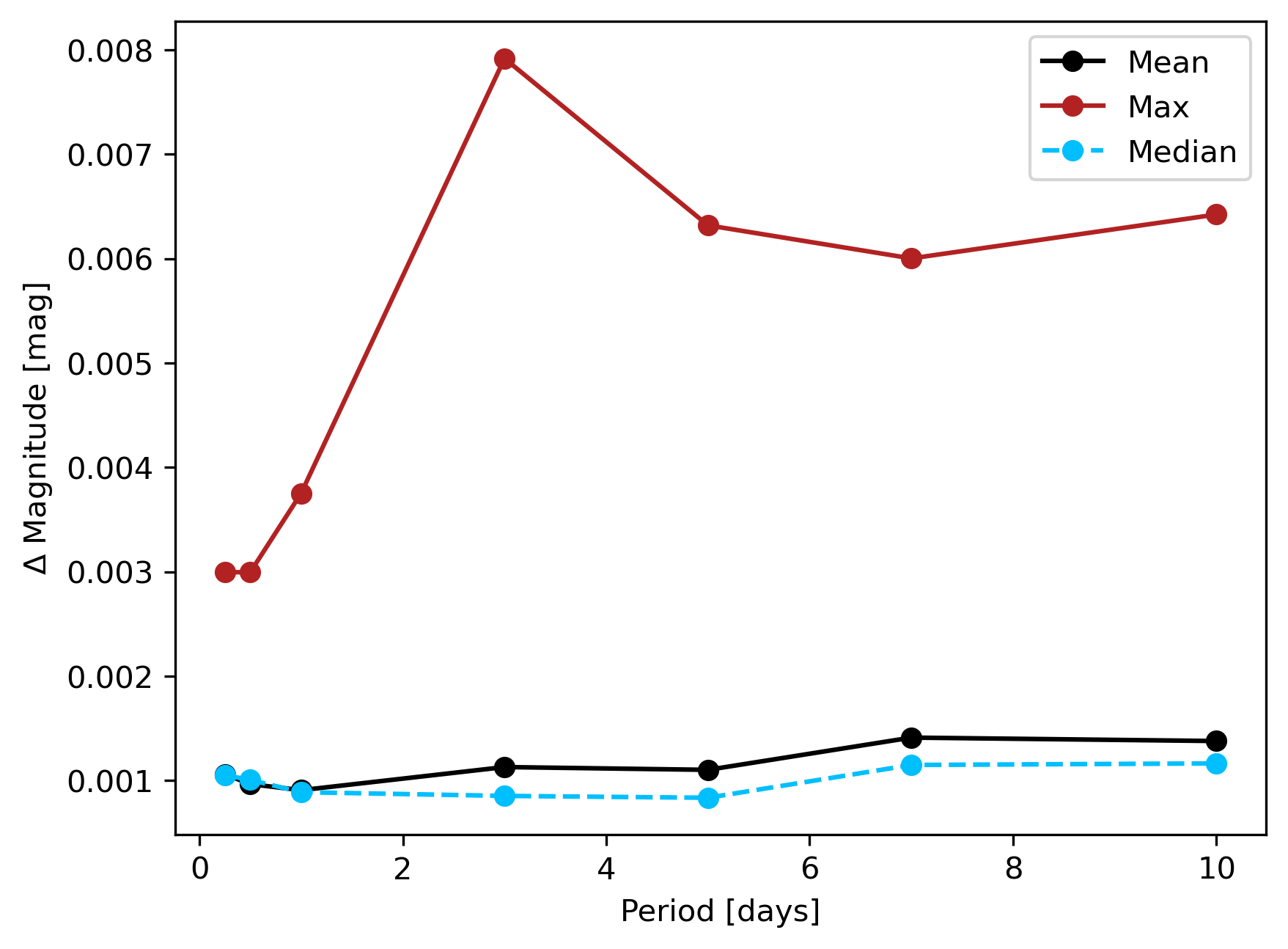}
    \caption{Same as Figure~\ref{fig:appendix_residuals_1000}, but for the medium-resolution mesh case (5120 mesh elements in \spice, 5806 in \phoebe). Residuals remain small and show little dependence on parameters other than mesh resolution.}
    \label{fig:appendix_residuals_5000}
\end{figure}

\begin{figure}
    \centering
    \includegraphics[width=\linewidth]{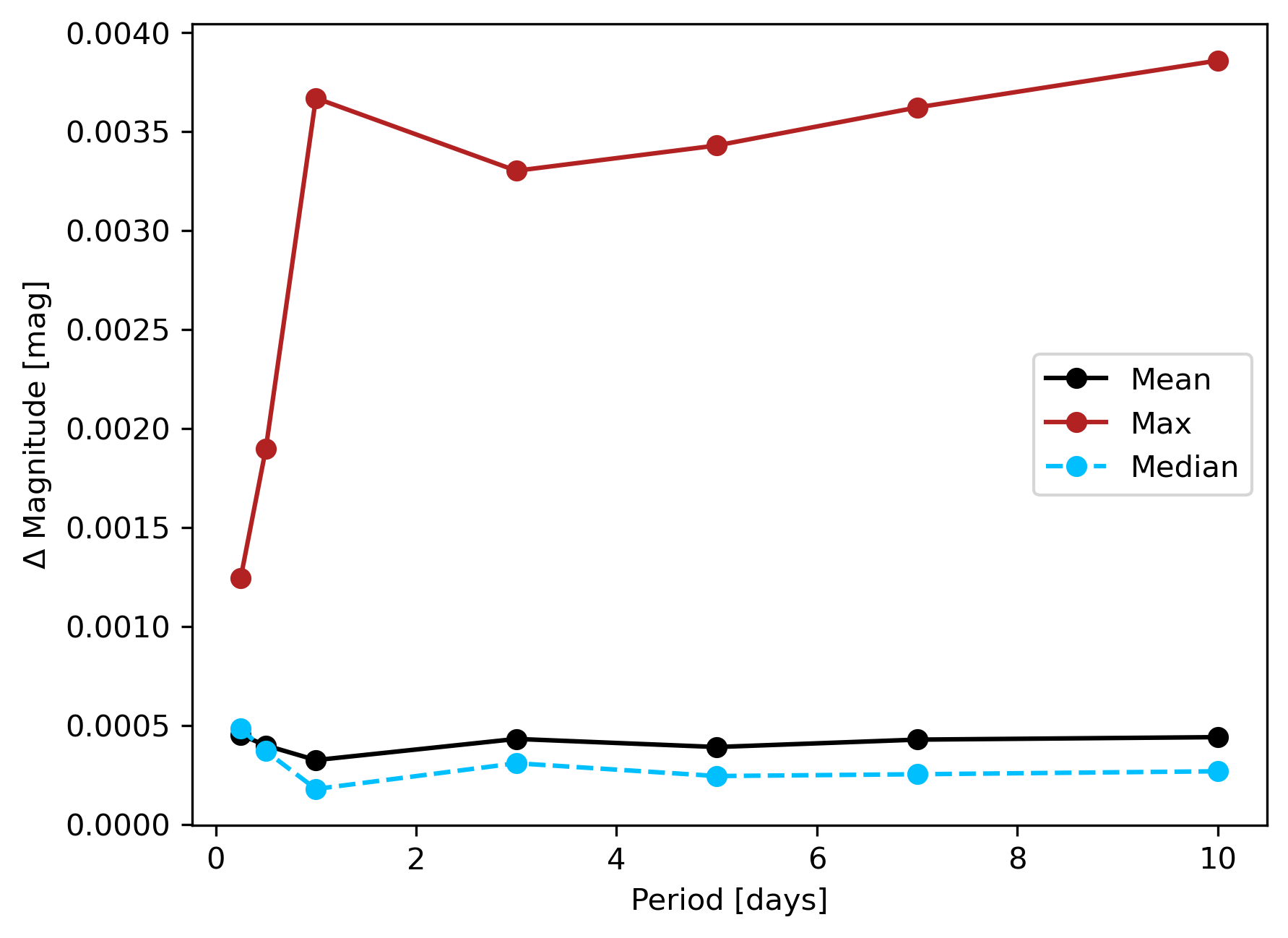}
    \caption{Same as Figure~\ref{fig:appendix_residuals_1000}, but for the high-resolution mesh case (20\,480 mesh elements in \spice, 22\,538 in \phoebe). Residuals are below typical ground-based photometric precision, indicating excellent agreement between the two codes at high mesh resolution.}
    \label{fig:appendix_residuals_20000}
\end{figure}

A more detailed examination of the differences between light curves generated with \texttt{PHOEBE} and \texttt{SPICE} is presented in Figures~\ref{fig:appendix_residuals_1000}, \ref{fig:appendix_residuals_5000}, and~\ref{fig:appendix_residuals_20000}. For clarity, only the dependence on period is shown. The results indicate that the residuals are primarily governed by the mesh resolution, with discrepancies remaining below the typical precision limits of ground-based surveys for high-resolution meshes (i.e., meshes with more than 20,000 elements). As the residuals exhibit minimal dependence on other parameters such as mass ratio or eccentricity, we conclude that the dominant source of variation arises from the different mesh construction schemes employed in \phoebe\ and \spice, which result in differing mesh element count for all cases.

\section{Code Optimizations}
\label{sec:appendix_code_optimizations}

The spectrum simulation code employs several memory and computation optimisation strategies, including chunking and JAX-specific features, to enable efficient handling of meshes with a large number of mesh elements.

\subsection{Chunked computation}

To efficiently handle high-resolution meshes and extensive wavelength grids, we implement a chunked computation strategy for calculating synthetic stellar flux. This approach processes the stellar surface in manageable segments rather than attempting to compute the entire surface at once. Following the radiative transfer formulation in Equation \ref{eq:synthetic_flux}, our chunked computational approach can be mathematically expressed as:
\begin{equation}
F_\lambda = \sum_{j=1}^{N_{\text{chunks}}} \left( \sum_{i \in \mathrm{chunk}_j} I_\lambda(\mu_i, \vec{p}_i) A_{\mathrm{proj},i} \frac{1}{d^2} \right),
\end{equation}
\noindent where index $i$ represents a mesh element within the currently processed chunk $j$, where chunk is internally represented by a subset of the whole mesh elements' set. The chunking strategy mitigates memory overflow issues that would otherwise occur when computing synthetic spectra for meshes with large numbers of elements. The approach introduces a configurable parameter - chunk size — that establishes a trade-off between memory utilisation and computational efficiency. No precision losses are introduced, as all elements are eventually processed. Smaller chunks reduce memory requirements at the cost of increased computation time, allowing users to optimise performance according to their specific hardware constraints and computational needs.

\subsection{JAX checkpoints}
The optimisation technique \texttt{@partial(jax.checkpoint, prevent\_cse=False)} applied to chunk processing functions represents a strategic performance trade-off. This "rematerialisation" decorator instructs JAX to recompute specific values during backpropagation rather than storing all intermediate results, significantly reducing memory requirements at the cost of additional computation time. By setting \texttt{prevent\_cse=False}, the code allows common subexpression elimination, letting JAX identify and optimise redundant calculations within the chunked operations. This combination of chunking and checkpointing enables the efficient processing of large spectral datasets that would otherwise exhaust available memory, while maintaining differentiability throughout the computational pipeline.

\section{Technical Validation}
\label{sec:appendix_technical_validation}

\subsection{Solar luminosity test}
\label{sec:technical_validation_solar_luminosity_test}

The solar luminosity is reproduced using a simple setup of a 20480-element mesh and the blackbody model with the same temperature as the solar surface value 5772~K \citep{iau}. The reason for adopting a black-body model is due to the limited wavelength coverage of the \texttt{TransformerPayne}, which prevents the recovery of the bolometric flux. 
Black-body intensities are calculated for a grid of 10,000 wavelength points evenly sampled between 1 $\mathring{A}$ and 100,000 $\mathring{A}$. Energy output at each surface area in all directions is calculated and summed for all mesh elements, following the procedure outlined in Appendix \ref{sec:appendix_radiative_quantities}. 
The result is then divided by a correction factor $S_\mathrm{ratio}=0.99970102$, which is the ratio between the used mesh surface area and the area of an exact sphere (see Appendix \ref{sec:appendix_mesh_construction}). The solar luminosity so derived is $3.824\cdot10^{33}$ erg/s, within $0.1$ per cent of the value expected from the Stefan-Boltzmann definition. 

\subsection{Solar constant test}

To further validate \spice, we compute the solar irradiance at 1 AU. Using the same solar model and wavelength range as in Section \ref{sec:technical_validation_solar_luminosity_test}, we evaluate the emergent specific intensity toward a distant observer and integrate it over wavelength to obtain the total irradiance (in $\mathrm{kW\,m^{-2}}$). This differs from the luminosity test in Section \ref{sec:technical_validation_solar_luminosity_test}, which integrates intensity over all outgoing angles on the stellar surface to recover the bolometric luminosity; here we use only the line-of-sight emergent intensity appropriate for a distant observer. After correcting for the small surface-area offset of the mesh model, the result is $1.360 \mathrm{kW\,m^{-2}}$, consistent with the measured solar constant \citep{solar_constant}.

\subsection{Blackbody Pulsation}
\label{sec:appendix_blackbody_pulsation}

\begin{figure}
    \centering
    \includegraphics[width=\linewidth]{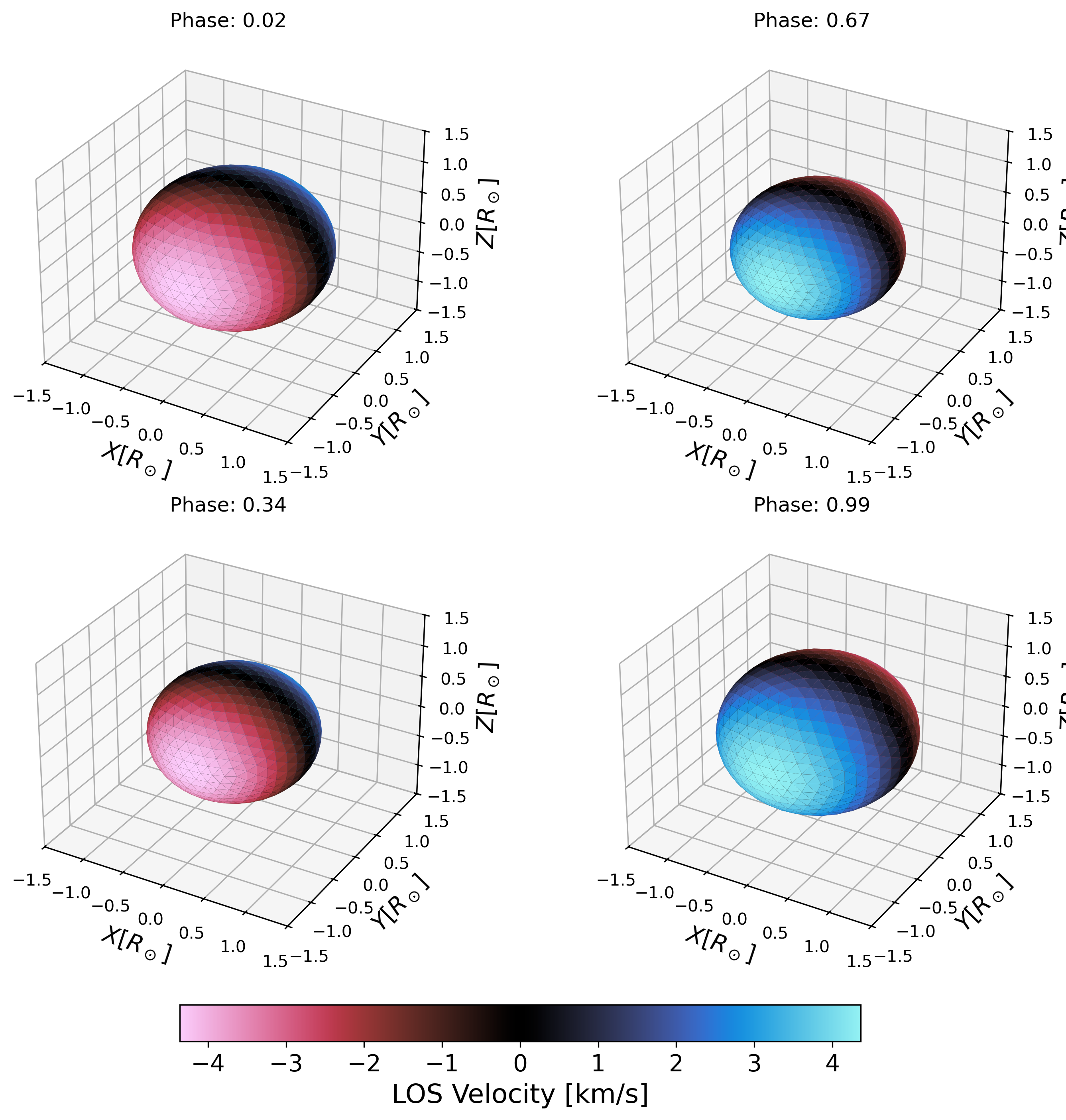}
    \caption{Line-of-sight velocities for a radially pulsating mesh model evaluated at several pulsation phases. The example uses a blackbody atmosphere with a single radial mode ($l = m = 0$), period $P = 1$~d, and maximum radial displacement $A_\text{max} = 0.1\,R_\text{star}$.}
    \label{fig:appendix_pulsation_los_phase_examples}
\end{figure}

A simple test for verifying the implementation of pulsations is a black-body model atmosphere of a star with only radial pulsations of significant amplitude. In our case, we added radial pulsation (or correspondingly, spherical harmonic function pulsation with a single mode of orders $m=0$ and $l=0$) of period $P=1\ d$ and maximum amplitude $A_\text{max}=0.1\ R_\text{star}$ to a star with surface temperature $T=5772\ K$ and radius of $1\ R_\odot$. Line-of-sight velocities for the pulsating mesh model for a few phases are illustrated in Figure \ref{fig:appendix_pulsation_los_phase_examples}.

\begin{figure}
    \centering
    \includegraphics[width=\linewidth]{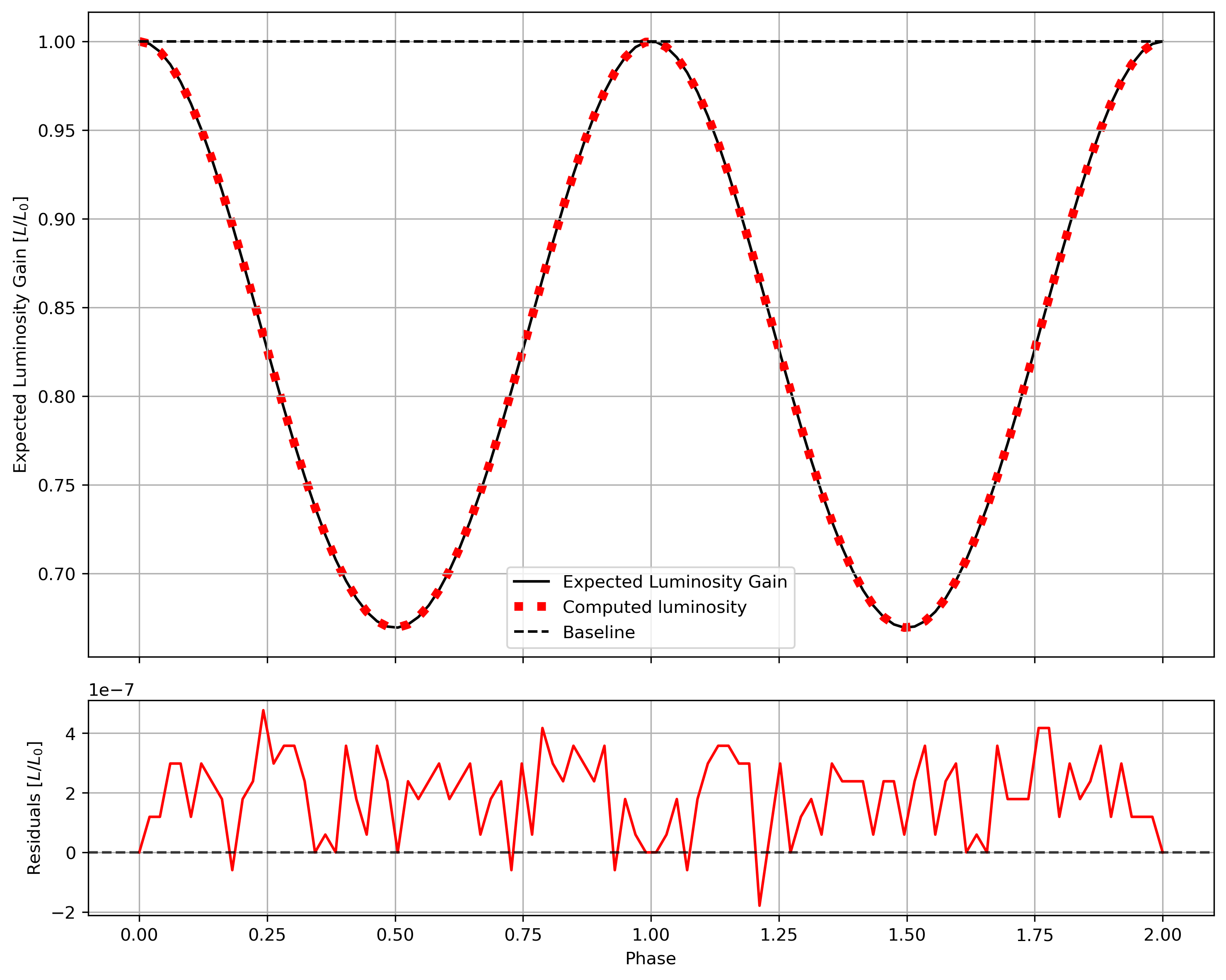}
    \caption{Expected luminosity gain due to the increase in stellar surface area from radial pulsations compared with the simulated observed flux for the pulsating mesh shown in Figure~\ref{fig:appendix_pulsation_los_phase_examples}. The excellent agreement (at the level of $10^{-7}$) validates the pulsation implementation in \spice.}
\label{fig:appendix_pulsation_luminosity_gain}
\end{figure}

For this simplified test case, we have not introduced any temperature change due to the radius change, although SPICE can readily implement this with a mesh parameter modification. In this toy model, luminosity changes are thus due to the change in surface area of the pulsating mesh model with 1,280 mesh elements. We have compared the simulated luminosity gain—computed by integrating the emergent intensities over the stellar mesh—to the analytic gain expected from the radius increase, and find agreement at the level of $10^{-7}$. The results of this comparison are demonstrated in Figure \ref{fig:appendix_pulsation_luminosity_gain}. These changes can also be seen in simulated passband luminosities. In the example studied here, with no surface temperature variations, the only effect contributing to the colour changes are Doppler shifts, which is however negligible due to the small radial velocity of order a few km/s.\

\subsection{Velocity effects}
\label{sec:appendix_velocity_effects}

\begin{figure*}
    \centering
    \includegraphics[width=0.9\textwidth]{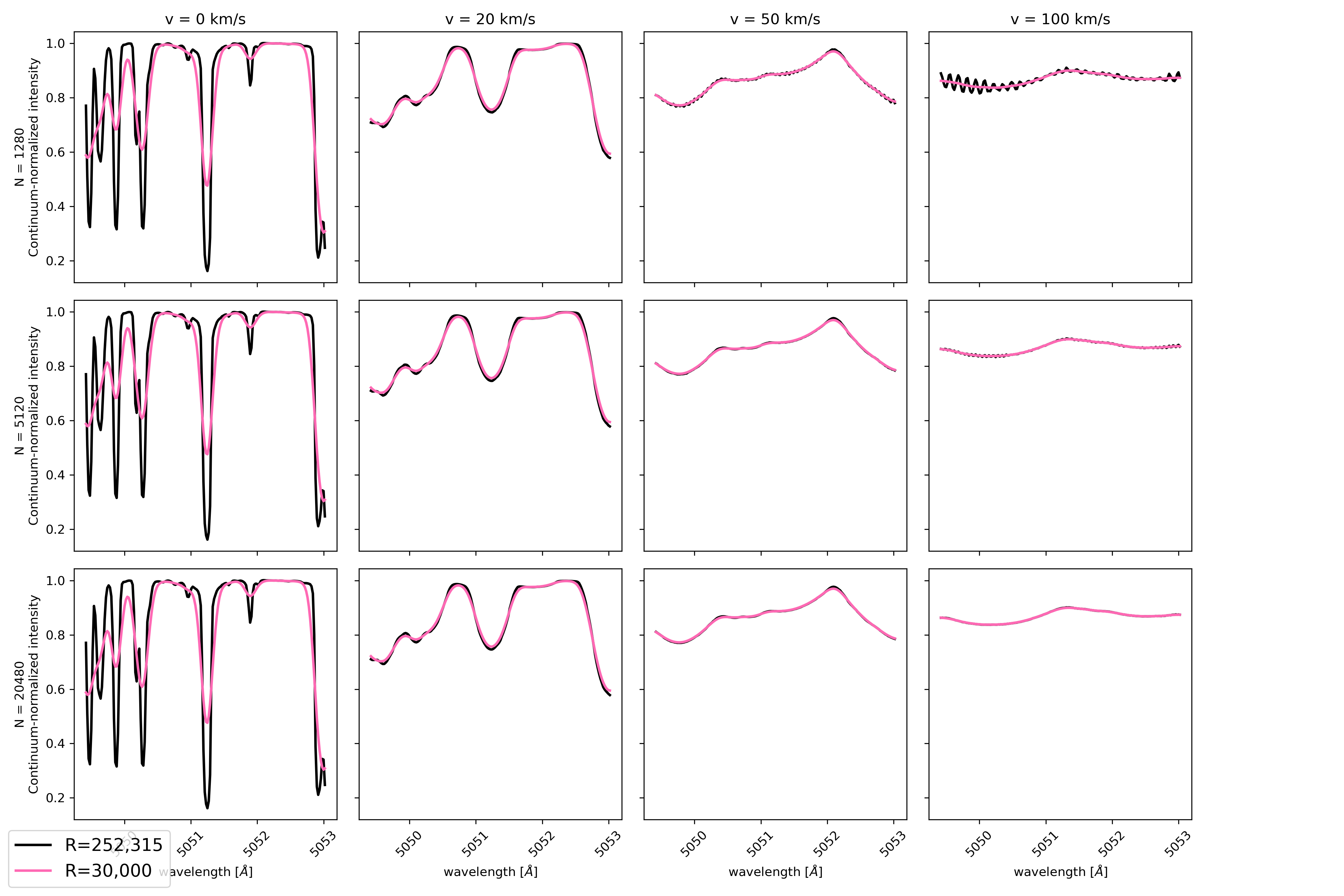}
    \caption{Line-profile variations caused by discretisation of the velocity field for meshes with different resolutions and equatorial rotational velocities. Each row corresponds to a different number of mesh elements $N$, and each column to a different equatorial velocity $v$. Spectra are shown for two spectral resolutions, demonstrating that mesh resolution has a negligible impact at lower resolving power, but can introduce small artefacts at high resolution for rapidly rotating stars if the mesh is too coarse.}
    \label{fig:appendix_velocity_effects_in_lines}
\end{figure*}

\begin{table}
\centering
\caption{Relative differences (\%) for different spectral resolution values $R$, mesh resolutions (quantified by the number of mesh elements), and equatorial rotation velocities v in km/s.}
\begin{tabular}{c|cccc}
\hline
\textbf{Mesh Resolution} & \textbf{v=0} & \textbf{v=20} & \textbf{v=50} & \textbf{v=100} \\
\hline
\multicolumn{5}{c}{\textbf{R = 252,315}} \\
\hline
N=1280 & 0.08197 & 0.16673 & 0.65867 & 3.61694 \\
N=5120 & 0.01631 & 0.12971 & 0.09428 & 0.45294 \\
\hline
\multicolumn{5}{c}{\textbf{R = 70,000}} \\
N=1280 & 0.04665 & 0.07297 & 0.08725 & 1.39705 \\
N=5120 & 0.00932 & 0.05783 & 0.04911 & 0.03631 \\
\hline
\multicolumn{5}{c}{\textbf{R = 30,000}} \\
\hline
N=1280 & 0.01124 & 0.05158 & 0.06611 & 0.16404 \\
N=5120 & 0.00224 & 0.02249 & 0.01769 & 0.02328 \\
\hline
\end{tabular}
\label{tab:line_effect_percentages}
\end{table}

The mesh density directly affects the accuracy of spectral line-profile calculations, especially for rapidly rotating stars where discretisation of the velocity field becomes critical. As shown in Figure \ref{fig:appendix_velocity_effects_in_lines}, increasing the mesh resolution from N = 1280 (top row) to N = 20480 (bottom row) reveals systematic differences in the synthesized Fe I (5049.82 {\r A}) line profiles across rotational velocities of 20, 50, and 100 km s$^{-1}$ at spectral resolution R = 252,315.  
At lower velocities (v = 20 km s$^{-1}$), even coarse meshes capture the key line-profile features well, whereas at higher velocities (v = 100 km s$^{-1}$), low-resolution models exhibit fine-scale oscillations caused by insufficient sampling of the velocity-field gradients across the visible stellar disk when studying precision spectroscopy. This effect diminishes with decreasing spectral resolution and becomes negligible for low-resolution spectra.  

We quantified the resulting noise as the maximum residual between spectra from lower- and highest-resolution models (20480 elements), summarised in the accompanying table \ref{tab:line_effect_percentages}. Even in the most extreme case, the offset remains 3.65\%. We therefore recommend using sufficiently dense meshes for stars with rapid rotation or pulsations to prevent numerical artefacts. Adaptive mesh-refinement methods will be included in future releases to improve both efficiency and accuracy in critical regions such as equatorial zones, which experience the largest line-of-sight velocities.


\bsp	
\label{lastpage}
\end{document}